\documentclass[twocolumn,showpacs,pra]{revtex4}
\usepackage{amsmath, amsthm, amssymb}
\usepackage{graphicx}
\usepackage{dcolumn}
\usepackage{bm}
\usepackage{epsf}
\usepackage{amsfonts, bm}
\usepackage{amsmath}
\usepackage{amssymb}
\usepackage[latin2]{inputenc}
\usepackage[T1]{fontenc}
\usepackage{epsfig,graphicx}
\usepackage{mathrsfs}
\usepackage{verbatim}

\newcommand{\1}{{\bf 1}}

\newcommand{\dep}{\mathcal{D}}

\newcommand{\tr}{\text{tr}}

\newcommand{\de}{\text{d}}

\newcommand{\mh}{\mathcal{H}}
\newcommand{\V}{\bm F}

\def\be{\begin{equation}}
\def\ee{\end{equation}}
\def\bea{\begin{eqnarray}}
\def\eea{\end{eqnarray}}
\def\bma{\begin{mathletters}}
\def\ema{\end{mathletters}}

\def\C{\hbox{$\mit I$\kern-.7em $\mit C$}}

\newcommand{\ket}[1]{|#1\rangle}

\newcommand{\ketbra}[2]{|#1\rangle \langle #2|}
\newcommand{\proj}[1]{|#1\rangle \langle #1|}
\newcommand{\ev}[1]{\langle #1 \rangle}

\def\one{\leavevmode\hbox{\small1\normalsize\kern-.33em1}}

\newlength{\dinwidth}
\newlength{\dinmargin}
\DeclareMathAlphabet{\scr}{U}{rsfs}{m}{n}

\begin{document}
\renewcommand\thefootnote{\fnsymbol{footnote}}

\title{Structural approximations to positive maps and entanglement breaking channels}

\author{J. K. Korbicz$^{1,2,3,4}$\footnote{jkorbicz@mif.pg.gda.pl}, M. L. Almeida$^{1}$, J. Bae$^{5}$, M. Lewenstein$^{1,6}$ and A. Ac\'\i n$^{1,6}$}

\affiliation{$^1$ ICFO--Institut de Ci\`{e}ncies Fot\`{o}niques,
Mediterranean Technology Park, 08860 Castelldefels (Barcelona),
Spain\\
$^2$ Dept. d'Estructura i
Constituents de la Mat\`eria, University of Barcelona, 08028 Barcelona, Spain\\
$^3$ Faculty of Applied Physics
and Mathematics, Technical University of Gda\'{n}sk, 80-952
Gda\'{n}sk, Poland\\
$^4$National Quantum Information Centre of
Gda\'{n}sk, 81-824 Sopot, Poland\\
$^5$ School of Computational Sciences, Korea Institute for Advanced Study, Seoul 130-012, Korea\\
$^6$ ICREA-Instituci\'o Catalana de Recerca i Estudis Avan\c cats,
08010 Barcelona, Spain }

\date{\today}

\setcounter{footnote}{0}
\renewcommand\thefootnote{\arabic{footnote}}

\begin{abstract}
Structural approximations to positive, but not completely positive
maps are approximate physical realizations of these non-physical
maps. They find applications in the design of direct entanglement
detection methods. We show that many of these approximations, in the
relevant case of optimal positive maps, define an entanglement
breaking channel and, consequently, can be implemented via a
measurement and state-preparation protocol. We also show how our
findings can be useful for the design of better and simpler direct
entanglement detection methods.
\end{abstract}

\maketitle

\section{Introduction}
Entanglement is one of the most important, and presumably necessary,
ingredients of quantum information processing \cite{review-horo}.
For this reason there is a considerable interest both in theory and
experiments in designing feasible and efficient ways of entanglement
detection. %Particularly important in this respect are entanglement
%detection schemes which require only local measurements of the
%entangled parts of the composite system.
Indeed, there has been a
lot of progress in this problem recently. The most frequently used and
investigated entanglement detection methods include: i) tomography of the quantum state with
local measurements, useful for low dimensional systems provided
entanglement criteria for the states in question are known
\cite{white98,blatt1,blatt2}, but impractical for higher dimensional
systems; ii) methods based on detecting only some elements of the
density matrix for a continuous family of measuring devices
settings, such as the method of entanglement visibility
\cite{visibility}; iv) tests of generalized Bell inequalities
\cite{Bell}, although there are states that despite
being entangled do not violate any Bell inequality \cite{Werner,lhv} nor
any known Bell inequality \cite{reviewbell}; v) entanglement
witnesses \cite{horo96,terhal};  vi) direct entanglement detection schemes, for pure \cite{huelga} or mixed states \cite{Mintert} and, in particular, using structural approximations to positive maps \cite{pawel,pawel_ekert}; vii) ``nonlinear'' entanglement witnesses \cite{guehne1} and viii) methods employing measurements of variances \cite{guehne2} or even higher order
correlation functions \cite{korbicz,blatt2}, or relying on entropic
uncertainty relations \cite{guehne3}. The methods v) and vi) are the
subject of the present paper and we discuss them in more detail
below. First we recall some basic definitions.

\paragraph{\bf Entanglement Witnesses.}
An observable $E=E^{\dag}$ is called an \emph{entanglement witness}
if and only if, for all separable states $\sigma$, the average
$\tr(E\sigma)\ge 0$ and there exists an entangled state $\varrho$
for which $\tr(E\varrho)<0$. As shown in Ref. \cite{horo96},
the Hahn-Banach theorem implies that for every entangled state
$\varrho$, there exists a witness $E$ that detects it, i.e.
$\tr(E\varrho)<0$. Conversely, the state $\sigma$ is separable if
and only if for all witnesses it holds $\tr(E\sigma)\ge 0$. As has
been pointed out in Ref. \cite{guehnepra}, entanglement witnesses
can be efficiently measured with local measurements and, more
importantly, one can optimize the complexity of this measurement
with respect to, for instance, the number of measuring device
settings. Nowadays, entanglement witnesses are routinely used in
experiments to detect entanglement in bipartite \cite{demartini} and
multipartite \cite{blatt1,harald} systems.

\paragraph{\bf Positive Maps.}
A related concept is that of a \emph{positive map}. Let
$\mathcal{B}(\mh_A)$ and $\mathcal{B}(\mh_B)$ denote the spaces of
bounded operators on Hilbert spaces $\mh_A$ and $\mh_B$
respectively. Then a linear map $\Lambda\colon \mathcal{B}(\mh_A)\to
\mathcal{B}(\mh_B)$ is called positive if $\Lambda(\varrho)\ge 0$
for every $\varrho\ge 0$. However, not every positive map can be
regarded as physical, describing e.g. a quantum channel or the reduced
dynamics of an open system: a stronger positivity condition is
required \cite{Kraus}. Namely, a map $\Lambda$ is physical whenever it is
\emph{completely positive}, which means that the extended map
$\1\otimes\Lambda\colon \mathcal{B}(\mathcal{K}\otimes\mh_A) \to
\mathcal{B}(\mathcal{K}\otimes\mh_B)$ is positive for any extension
$\mathcal K$.

Again, as shown in Ref. \cite{horo96} (see also \cite{Woronowicz}),
a state $\varrho\in\mathcal B(\mh_A\otimes\mh_B)$ is entangled if
and only if there exists a positive, not completely positive map
$\Lambda\colon\mathcal B(\mh_B)\to\mathcal B(\mh_A)$ that detects
$\varrho$, i.e. $[\1\otimes\Lambda](\varrho)$ is not positive
definite. A paradigm example of a positive but not completely
positive map is transposition, $T$, whose great significance for
separability was first realized in Ref. \cite{PPT}. It turns out to
detect all the entangled states in $\mathcal B(\mathbb C^2\otimes
\mathbb C^2)$ and
 $\mathcal B(\mathbb C^2\otimes \mathbb C^3)$ \cite{horo96}. However, as
it is well known \cite{be} (see also e.g. Ref. \cite{review-horo}
and references therein), in higher dimensions there are entangled
states which possess the positive partial transpose (PPT) property.

Entanglement witnesses and positive maps \cite{notepos} are related
through the Jamio\l kowski isomorphism \cite{jam}. Let $E\in
\mathcal B(\mh_A\otimes\mh_B)$. From this moment on we assume that
the considered Hilbert spaces are finite dimensional,
$\text{dim}\mh_{A,B}=d_{A,B}<\infty$ (for an example of infinite
dimensional generalization of Jamio\l kowski isomorphism see Ref.
\cite{cmp}). We then define $\Lambda_E\colon \mathcal B(\mh_A) \to
\mathcal B(\mh_B)$ as follows:
\begin{equation}\label{LE}
\Lambda_E(\varrho)=d_A\tr_A\left[E(\varrho^T\otimes\1)\right].
\end{equation}
Conversely, introducing a maximally entangled vector in
$\mh_A\otimes\mh_A$,
\begin{equation}\label{P+}
\ket{\Phi_+}=\frac{1}{\sqrt{d_A}}\sum_{i=1}^{d_A}|ii\rangle,
\quad P_+=|\Phi_+\rangle\langle\Phi_+|
\end{equation}
we define for each map $\Lambda\colon \mathcal B(\mh_A) \to \mathcal
B(\mh_B)$ an operator
\begin{equation}\label{EL}
E_\Lambda=\1\otimes\Lambda(P_+)
\end{equation}
acting on $\mathcal B(\mh_A\otimes\mh_B)$. Then $\Lambda$ is
positive if and only if $E_\Lambda$ is an entanglement witness,
since $E_{\Lambda_E}=E$ \cite{jam}. Moreover,  $\Lambda$ is
completely positive if and only if $E_\Lambda\ge 0$, i.e.
$E_\Lambda$ is a (possibly unnormalized) state.

\paragraph{\bf Structural Physical Approximation.}
Positive maps are stronger detectors of entanglement than the
corresponding witnesses, despite the described Jamio\l kowski
isomorphism. They detect the same states as the corresponding
witnesses plus those obtained through local invertible
transformation on one side. Unfortunately generic positive maps are
not physical and their action cannot be directly implemented. It is
therefore challenging to try to find a physical way to approximate
the action of a positive map. This is the goal of the
\emph{structural physical approximation}
\cite{pawel,pawel_ekert}. The idea is to mix a positive map
$\Lambda$ with some simple completely positive map (CPM), making the
mixture $\tilde \Lambda$ completely positive. The resulting map can then be realized in the laboratory and its action characterizes
entanglement of the states detected by $\Lambda$. In the particular
example studied in Refs.~\cite{pawel,pawel_ekert} this idea has been
applied to the map $\Lambda=\1\otimes T$. Although experimentally
viable, this method is not easy to implement since, at least in its
original version, it requires highly nonlocal measurements: subsequent
applications of $\tilde \Lambda$ followed by optimal spectrum
estimation. A more detailed discussion on this entanglement detection scheme is given in Section \ref{entdet}.

In the case of finite dimensional Hilbert spaces, we can, without
lost of generality, restrict our attention to {\it contractive}
structural approximations, i.e. ${\rm tr}(\tilde\Lambda(\varrho))\le
1$, for $\tr(\varrho)= 1$. If the initial map is not contractive, we
can always define
$\tilde\Lambda'(\varrho)=\tilde\Lambda(\varrho)/{\rm
tr}(\tilde\Lambda(\varrho^*))$, where ${\rm
tr}(\tilde\Lambda(\varrho^*))=\max_{\varrho: \tr(\varrho)= 1}{\rm
tr}(\tilde\Lambda(\varrho))$) and the maximum is attained for some
$\varrho^*$ due to the compactness of the set of all states.
Contractive CPM's can be realized probabilistically as a partial
result of a generalized measurement (see Ref. \cite{pawel}).

\paragraph{\bf Entanglement Breaking Channels.}
A different use of maps in the context of quantum theory concerns the
description of \emph{quantum channels}, which are completely
positive and trace-preserving maps. A channel $\Lambda$ that
transforms any state $\varrho$ into a separable state $\Lambda(\varrho)$
is called \emph{entanglement breaking} (EB) \cite{ruskai}. Clearly, these channels are useless for entanglement distribution. In Ref.~\cite{ruskai}, the following equivalence was obtained:
\begin{enumerate}
    \item The channel $\Lambda$ is EB.
    \item The corresponding state $E_\Lambda$ is separable.
    \item The channel can be represented in the Holevo form:
\begin{equation}\label{holevo}
\Lambda(\varrho)=\sum_k\tr(F_k\varrho)\varrho_k,
\end{equation}
with some positive operators $F_k\geq 0$ defining a generalized
measurement \cite{peresbook}, $\sum_k F_k=\1$, and states
$\varrho_k$ determined only by $\Lambda$.
   \end{enumerate}
The last property above means that the action of an EB channel can
be substituted by a measurement and state-preparation protocol.
Moreover, from the separable decomposition of the state $E_\Lambda$
\begin{equation}\label{sep dec}
E_{\Lambda}=\sum_kp_k\proj{v_k}\otimes\proj{w_k}
\end{equation}
with $\ket{v_k}\in \mathcal H_A$ and $\ket{w_k} \in \mathcal H_B$,
one obtains the following explicit Holevo representation of
$\Lambda$ \cite{ruskai}:
\begin{equation}\label{holevo_expl}
\Lambda(\varrho)=\sum_k
\proj{w_k}\tr\Big[\left(d_Ap_k\proj{\bar{v_k}}\right)\varrho\Big],
\end{equation}
where the overbar denotes the complex conjugation. The positive
operators $\{d_Ap_k\proj{\bar v_k}\}$ define a properly normalized
measurement due to the trace-preserving property of $\Lambda$.

Notice that these results can easily be extended to the case of
contracting maps. Then, a Holevo decomposition is still possible for
EB maps with the positive operators $F_k$ defining a partial
measurement, $\sum_k F_k<\1$.

In this paper we address the question of implementation  of
structural approximations to positive maps through (generalized)
measurements. In particular, we study structural approximations to
maps $\Lambda\colon\mathcal B(\mh_A)\to\mathcal B(\mh_B)$ obtained
through minimal admixing of white noise:
\begin{equation}\label{strapprox}
\widetilde\Lambda(\rho)=p\,\tr(\rho)\frac{\bf
1}{d_B}+(1-p)\Lambda(\rho).
\end{equation}
Minimal means here that we take the smallest noise probability
$0<p<1$ for which $\tilde\Lambda$ becomes completely positive. Now
the key question is when such $\tilde\Lambda$ can be implemented
through generalized measurements, i.e. when they correspond to EB
maps according to Eq.~\eqref{holevo}. As a consequence, we are led
to study  the separability of witnesses of the form
\begin{equation}\label{genform}
\widetilde E_{\Lambda}={\bf 1}\otimes\widetilde\Lambda(P_+)=
\frac{p}{d_Ad_B}{\bf 1}+ (1-p)E_\Lambda
\end{equation}
for minimal $p$ such that
\begin{equation}\label{CP}
\widetilde E_{\Lambda}\ge 0.
\end{equation}
Recall that this is equivalent to $\widetilde\Lambda$ being
completely positive.
%More generally, we will sometimes consider also SA's in whihc we will replace $I_{AC}$,
% by $\tilde I_{AC}=G_A\otimes G_C \hat I_{AC}G^{\dag}_A\otimes G^{\dag}_C$,
% where $G$'s are invertible. We will call these types of
%SA's simple SA's (SSA).
In general, we will consider contractive maps $\tilde\Lambda$ and
ask whether they correspond to EB maps, not necessarily
trace-preserving. Note that if a CPM $\tilde\Lambda$ is contractive
and EB, then there exists an EB extension to a trace-preserving map,
$\tilde\Lambda'(\varrho)=\tilde\Lambda(\varrho) + [{\rm
tr}(\varrho)-{\rm tr}(\tilde\Lambda(\varrho))]{\bf 1}/d$. In the
language of witnesses, trace preservation means ${\rm
tr}_BE_{\tilde\Lambda}={\bf 1}_A$.

The main subject of the paper is the following conjecture: \\

\noindent{\bf Conjecture:} Structural physical approximations to
{\it optimal} positive  maps correspond to entanglement breaking
maps. Equivalently, structural physical approximations to {\it
optimal} entanglement witnesses $E$
are given by (possibly unnormalized) separable states. \\

We prove the above conjecture in several special cases and discuss a
large number of generic examples providing evidence for its
validity. This is done for both decomposable entanglement witnesses,
that detect only entangled states with negative partial
transposition (NPT), and also for non-decomposable witnesses which
in addition detect PPT entangled states. Once more, we expect the
conjecture to be valid in general, but in some cases we restrict our
examples to structural approximations which are trace-preserving.
Note that such restriction makes the conjecture weaker, since every
contractive EB channel has a trace preserving EB extension, but not
{\it vice-versa}.

The importance of our result is twofold: i) if the conjecture is
true, structural physical approximations to optimal maps admit a
particularly simple experimental realization---they correspond to
generalized measurements \cite{remik}; ii) the results shed light on
the geometry of the set of entangled and separable states (cf. Ref.
\cite{karol}).

The paper is organized as follows. In Section \ref{optimal} we
recall the notions of decomposable and non-decomposable entanglement
witnesses and their optimality, based on the Refs. \cite{opt1,opt2}.
In Section \ref{dec} we concentrate on decomposable maps. First we
study  dimensions $2\otimes 2$ and $2\otimes 3$, where the
positivity of the partial transpose provides the separability
criterion. Here we show that in general, without the assumption of
optimality, the conjecture is not true, while it obviously holds for
optimal decomposable witnesses. Next, we discuss general
decomposable maps in $2\otimes 4$ systems, which are nontrivial due
to the existence of PPT entangled states. Other examples of maps in
$3\otimes 3$ systems satisfying the conjecture are presented in
Appendix B. We conclude this section proving the conjecture for the
transposition and reduction map \cite{reduction} in arbitrary
dimension.
%as well as an example which illustrates  why the
%conjecture is not true  without the assumption of simplicity.
Section \ref{nondec} is devoted to non-decomposable positive maps.
We start the discussion by analyzing the case of Choi's map, one of
the first examples of a map in this class. Then, we study a positive
map based on unextendible product bases (UPB's) \cite{terhal_upb}.
Finally, we end this section with an analysis of the Breuer-Hall map
\cite{Breuer,Hall}, which can be understood as the non-decomposable
version of the reduction criterion. Here symmetry methods turn out
to be indispensable. We introduce and study in some detail a new
family of states---unitary symplectic invariant states. The most
technical details of these states are mainly given in Appendix C,
where, as a byproduct, we show that this family includes also bound
entangled states. Finally, we study the physical approximation to
partial transposition, as this map is used in the direct
entanglement detection method proposed in \cite{pawel_ekert}. In the
latter case the analysis is again made possible due to symmetry
arguments, in particular the unitary $U\bar U VV$ symmetry (cf.
Refs. \cite{werner_sym,masanes}). The paper ends with the
conclusions in Section \ref{concl}.

\section{Optimality of Positive Maps and Entanglement Witnesses}\label{optimal}
The notion of optimality of positive maps and entanglement witnesses
has been introduced in Refs. \cite{opt1,opt2}. We review it here
without proofs, which can be found in the original papers. There are
two concepts of optimality: one general, and one strictly related to
non-decomposable positive maps (or entanglement witnesses) and PPT
entangled states. We focus below on entanglement witnesses---the
translation to positive maps is straightforward using the Jamio\l
kowski's isomorphism (cf. Eqs. (\ref{LE}) and (\ref{EL})).

\subsection{General Optimality}\label{GO}
Let us introduce the notion of general optimality first.
Given an entanglement witness $E$ we define:

\begin{itemize}

\item  $D_E=\{\varrho\ge 0 : \tr(E\varrho)<0\}$---the set
of operators detected by $E$.

\item \emph{Finer witness} --- given two
witnesses $E_1$ and $E_2$ we say that $E_2$ is
{\it finer} than $E_1$, if $D_{E_1}\subset D_{E_2}$, i.e. if all the
operators detected by $E_1$ are also detected by $E_2$.

\item \emph{Optimal witness} --- $E$ is optimal if
there exists no other witness which is finer than $E$.

\item $P_E=\{u\otimes v \in \mh_A\otimes\mh_B : \langle
u\otimes v|E u\otimes v\rangle=0\}$ --- the set of product vectors
on which $E$ vanishes. As we will show, these vectors are closely
related to the optimality property.

\end{itemize}
Vectors in $P_E$ play an important role regarding entanglement. A
full characterization of optimal witnesses is provided by the
following theorem:

{\bf Theorem 1:} A witness $E$ is optimal if and only if for all
operators  $P\ge 0$ and numbers $\epsilon>0$, $E'=E-\epsilon P$
is not an entanglement witness.

In this paper we will use the following important corollary:

{\bf Corollary 2:} If the set $P_E$ spans the whole Hilbert space
$\mh_A\otimes\mh_B$,  then $E$ is optimal.

\subsection{Decomposable witnesses}\label{DW}
There exists a class of entanglement witnesses which is very simple
to characterize---{\it decomposable entanglement witnesses}
\cite{Woronowicz}. Those are the witnesses which can be written in
the form: \be \label{DEW} E=Q_1+ Q_2^\Gamma, \ee where $Q_{1,2}\ge
0$ and $\Gamma$ refers to partial transposition with respect to the
second subsystem:
\begin{equation}
Q^\Gamma=({\bf 1}\otimes T)Q.
\end{equation}
As it is well known, these witnesses cannot detect PPT entangled
states. We recall here some simple properties of optimal
decomposable entanglement witnesses:

{\bf Theorem 2:} Let $E$ be a decomposable witness. If $E$ is optimal then it
can be written as $E=Q^\Gamma$, where $Q\ge 0$ contains no product
vector in its range.

This result can be slightly generalized as follows:

{\bf Theorem 2':} Let $E$ be a decomposable witness. If $E$ is optimal then
it can be written as $E=Q^\Gamma$, where $Q\ge 0$ and there is no
operator $P$ in the range of $Q$ such that $P^\Gamma\ge 0$.

\subsection{Non-decomposable witnesses}
Entanglement witnesses which are able to detect PPT entangled states
cannot be written in the form (\ref{DEW}) \cite{Woronowicz}, and are
therefore called {\it non-decomposable}. The present Subsection is
devoted to this kind of witnesses. The importance of
non-decomposable witnesses for detecting PPT entanglement is
reflected by the following:

{\bf Theorem 3:} An entanglement witness
is non--decomposable if and only if it detects
some PPT entangled state.

We now recall some definitions which are parallel to those provided
previously. Given a non-decomposable witness $E$, we define:

\begin{itemize}

\item $d_E=\{\varrho\ge 0: \varrho^\Gamma\ge 0 {\mbox{
and }} \tr(E\varrho)<0\}$ --- the set of PPT operators detected by
$E$.

\item \emph{Finer non-decomposable witness}---given two
non-decomposable witnesses $E_1$ and
$E_2$ we say that $E_2$ is {\it nd--finer} than $E_1$ if
$d_{E_1}\subset d_{E_2}$, i.e. if all PPT operators detected by
$E_1$ are also detected by $E_2$.

\item \emph{Optimal non-decomposable witness} --- $E$ is
optimal non-decomposable if there exists no other non-decomposable
witness which is nd-finer than $E$.

\end{itemize}

Again, vectors in $P_E$ play an important role regarding PPT
entangled states. The full characterization of optimal
non-decomposable witnesses is given by an analog of Theorem 1:

{\bf Theorem 4:} A non-decomposable entanglement witness $E$ is
optimal if and only if for all decomposable operators $D$ and $\epsilon>0$,
$E'= E-\epsilon D$ is not an entanglement witness.

Note that in principle non-decomposable optimality requires the
witness to be finer with regard to PPT entangled states only, so
that a  non-decomposable optimal witness does not have to be optimal
in the sense of Section \ref{GO}. However, this is not the case
since we have the following:

{\bf Theorem 5:} $E$ is an optimal non-decomposable entanglement
witness if and only if both $E$ and $E^\Gamma$ are optimal
witnesses.

and

{\bf Corollary 6:} $E$ is an optimal non-decomposable
witness if and only if $E^\Gamma$ is an optimal non-decomposable
witness.

In Ref. \cite{opt1} optimality conditions have been derived and
investigated for the case of $2\otimes N$-dimensional Hilbert
spaces. These conditions are, however, very complex and for the
purpose of the present work we will use Corollary 2 to check
optimality, even though it provides only a
sufficient condition.

\section{Decomposable maps}\label{dec}

This section is devoted to the study of the conjecture for
decomposable maps. We start by proving the conjecture for low
dimensional systems, namely $2\otimes 2$ and $2\otimes 3$. Then, we
provide some rather general results for $2\otimes 4$ systems.
Moreover, Appendix B contains several relevant examples of
decomposable maps in $3\otimes 3$ systems where the conjecture also
holds. Finally, we prove the conjecture in arbitrary dimension for
two of the most important examples of decomposable maps, the
transposition and reduction maps.

\subsection{$2\otimes 2$ and $2\otimes 3$}\label{2x2}
We begin with general examples in the lowest non-trivial dimensions.
Take $\Lambda$ to be a positive map from $\mathcal{B}(\mathbb C^2)$
to $\mathcal{B}(\mathbb C^2)$ or to $\mathcal{B}(\mathbb C^3)$.
Recall \cite{Woronowicz} that every such map is decomposable, i.e.
is of the form $\Lambda=\Lambda^{CP}_1+T\circ\Lambda^{CP}_2$ and
that its corresponding entanglement witness can be written as
(\ref{DEW}). We will first show that not every structural
approximation to $\Lambda$ is entanglement breaking. In other words,
the optimality of the positive map is essential for the conjecture.
For definiteness' sake we analyze the $2\otimes 2$-dimensional case,
but the argument also holds in $2\otimes 3$ systems.

Let us consider the entanglement witness $E_{\Lambda}$ corresponding
to $\Lambda$ and, $Q_1$ and $Q_2$ in \eqref{DEW} to be rank-one
operators of the form:
\begin{equation}
Q_2=\left[\begin{array}{cccc} a & 0 & 0 & a\\
                        0 & 0 & 0 & 0\\
                        0 & 0 & 0 & 0\\
                        a & 0 & 0 & a \end{array}\right], \quad
Q_1=\left[\begin{array}{cccc} 0 & 0 & 0 & 0\\
                            0 & b & b & 0\\
                            0 & b & b & 0\\
                            0 & 0 & 0 & 0 \end{array}\right]
\end{equation}
with real positive $a$ and $b$. Then the witness
\begin{equation}
Q_1+Q_2^{\Gamma}=\left[\begin{array}{cccc} a & 0 & 0 & 0\\
                                   0 & b & b+a & 0\\
                                   0 & b+a & b & 0\\
                                   0 & 0 & 0 & a \end{array}\right]
\end{equation}
is not positive and therefore $\Lambda$ is not completely positive.
From the general form (\ref{genform}) we obtain that:
\begin{eqnarray} \nonumber
 \widetilde E_{\Lambda}=&\frac{p}{4}{\bf 1}+
(1-p)\big(Q_1+Q_2^{\Gamma}\big)\\ \nonumber
               &=\left[\begin{array}{cccc}a'+c & 0 & 0 & 0\\
                                   0 & b'+c & b'+a' & 0\\
                                   0 & b'+a' & b'+c & 0\\
                                   0 & 0 & 0 & a'+c \end{array}\right],\label{cos}\\
a'=&(1-p)a, \quad b'=(1-p)b,\quad c=\frac{p}{4}.
\end{eqnarray}
This operator is positive for
\begin{equation}\label{chuj1}
p\ge \frac{4a}{4a+1},
\end{equation}
which is the condition for the structural approximation.

In order to study separability, it is enough to check the PPT condition,
as it is both a necessary and sufficient condition in the lowest
dimensions \cite{PPT}. Applying partial transposition we obtain that
the state (\ref{cos}) is not PPT, and hence entangled, for
\begin{equation}\label{chuj2}
p<\frac{4b}{4b+1}.
\end{equation}
Taking $b>a$ the condition (\ref{chuj1}) and (\ref{chuj2}) can be simultaneously
satisfied, thus giving a structural approximation which is not entanglement breaking.

Above we have considered a general positive map
from  $\mathcal B(\mathbb C^2)$ to $\mathcal B(\mathbb C^2)$
or to $\mathcal B(\mathbb C^3)$, i.e.
$\Lambda=\Lambda^{CP}_1+T\circ\Lambda^{CP}_2$.
Let us now consider an optimal one:
\begin{equation}\label{opt}
\Lambda=T\circ\Lambda^{CP},\quad E_\Lambda=Q^{\Gamma}.
\end{equation}
It immediately follows that any structural approximation to such
$\Lambda$ is entanglement breaking since
\begin{equation}\label{decPPT}
E_{\Lambda}^{\Gamma}=\frac{p}{4}{\bf 1}+(1-p)Q\ge 0,
\end{equation}
so that $E_{\Lambda}$ is separable. Thus, in the lowest dimensions
any structural approximation to an optimal map is entanglement
breaking.

More generally, for arbitrary dimension we immediately obtain from
Eq. (\ref{decPPT}) that any structural approximation to an optimal
decomposable map (\ref{opt}) (cf. Section \ref{GO}) gives rise to a
PPT state. However, in principle not necessarily that state is
separable, i.e. not necessarily $\widetilde\Lambda$ is entanglement
breaking, as PPT condition is no longer sufficient for separability
in higher dimensions.

\subsection{$2\otimes 4$}\label{2x4}
We now study optimal decomposable maps in $2\otimes 4$ dimensional
systems. The main characterization of such witnesses/maps is given
by Theorems 2 and 2' from Section \ref{DW} and we will use it
extensively in what follows. In some cases we will present general
results, while in others we consider what seem generic examples,
giving evidence supporting our conjecture. Other examples of
witnesses in $3\otimes 3$ systems fulfilling the conjecture are
presented in Appendix B.

Let us then consider systems with dimension $2\otimes 4$. There are
only three possibilities in this case, depending on the rank $r(Q)$
of the operator $Q$ (cf. Theorem 2, Section \ref{DW}): $r(Q)=1,2$,
or $3$. Higher ranks are not possible as then $Q$ would have a
product vector in its range and hence the witness $Q^\Gamma$ would
not be optimal~\cite{opt1}.

When $r(Q)=1$, then $Q$ is effectively supported in a $2\otimes 2$
subspace and the results of Section \ref{2x2} imply that its
structural approximation is entanglement breaking.

When $r(Q)=2$ there are two further possibilities: $Q$ is
supported either in a $2\otimes 3$ subspace or in the full
$2\otimes 4$ space. The first case is again covered by
Section \ref{2x2}. In the latter case, $Q$ can be written as
a sum of projectors:
\begin{equation}\label{2x4 r=2}
Q=P_\psi+P_\chi,\\
\end{equation}
where
\begin{eqnarray}
& &\psi=|0\rangle|f_1\rangle+|1\rangle|f_2\rangle,\\
& &\chi=|0\rangle|f_3\rangle+|1\rangle|f_4\rangle.
\end{eqnarray}
Here $\ket 0,\ket 1$ is the standard basis in $\mathbb C^2$ and
$f_1,\dots,f_4$ are vectors in $\mathbb C^4$. In the most general
case of contractive maps, $f_1$ and $f_2$ are orthogonal to $f_3$
and $f_4$, and this consists of the only condition required for the
proof. Note that if the vectors $f_i$ are mutually orthonormal, the
map is trace-preserving. Projectors in Eq.~\eqref{2x4 r=2} define a
decomposition of $\mathbb C^4$ into a direct sum $\mathbb C^2\oplus
\mathbb C^2$ and, hence, $Q$ has a block-diagonal form resulting
from a split $2\otimes (2\oplus 2)=(2\otimes 2)\oplus (2\otimes 2)$.
Applying the results of Section \ref{2x2} to each of the $2\otimes
2$-blocks we obtain the result.

We are left with the most interesting case: $r(Q)=3$. Take $P$ a
projector on the kernel of $Q$. The state $P$ has rank 5, is PPT and
possibly entangled. In this case we cannot prove the conjecture in
general, and we consider an example where the range of $P$ is
spanned by the product vectors $(1,\alpha)\otimes
(1,\alpha,\alpha^2,\alpha^3)$, for all complex numbers $\alpha$.
This can be always achieved applying a local invertible
transformation on $\mathbb C^4$ side (cf. \cite{soms}). Since, by
construction, $Q$ is supported on $P$'s kernel, it is supported on a
span of the vectors
\begin{eqnarray}
\label{2x4vectors}
& &|10\rangle-|01\rangle,\\
& &|02\rangle-|11\rangle,\\
& &|03\rangle-|12\rangle
\end{eqnarray}
where $|ij\rangle$ denotes the standard product basis of $\mathbb
C^2\otimes \mathbb C^4$.  As evidence to support our conjecture, one
can see that the structural approximation to $Q^\Gamma$, where $Q$
is given by the sum of projectors on the above vectors, is indeed
entanglement breaking. The details of the separability proof are
given in Appendix A. Note, that the EB map corresponding to
$Q^\Gamma$ is not trace-preserving, but as we mentioned in the
introduction it has an EB trace-preserving extension.

\subsection{Transposition}\label{trans} \label{ssec. trans}

We conclude the study of decomposable maps by proving the conjecture
in arbitrary dimension for two of the best known positive maps, the
transposition and reduction maps. Let us first consider structural
approximations to transposition $T\colon\mathcal B(\mh)\to \mathcal
B(\mh)$, $\mh\cong\mathbb C^d$, which is an optimal decomposable
map, for arbitrary dimension. The corresponding witness $\widetilde
E_{T}$, obtained from Eq. (\ref{genform}), turns out to be a Werner
state on $\mh \otimes \mh$:
%\begin{equation}\label{werner}
%\widetilde E_{T}={\bf 1}\otimes\widetilde T(P_+)
%=\frac{d(1-p)+p}{d^2}{\bf 1}
%-\frac{2(1-p)}{d}\Pi_-.
%\end{equation}
%Here $\Pi_-$ is the projector onto the skew-symmetric subspace
%$\mh\wedge \mh$.
\begin{equation}\label{werner}
\widetilde E_T=\frac{p}{d^2}\1 + \frac{1-p}{d}\V
\end{equation}
Here $\V$ is the flip operator, such that $\V \psi\otimes\phi=\phi\otimes\psi$.
One easily sees that $\widetilde E_{T}$ is positive, and hence
$\widetilde T$ completely positive, when
\begin{equation}\label{cond}
p\ge \frac{d}{d+1},
\end{equation}
which is the condition for the structural approximation for $T$.
To check the separability of $\widetilde E_{T}$, we use the fact
that the PPT criterion is necessary and sufficient for Werner
states. Since, for all $p$, we have
\begin{equation}
\widetilde E_{T}^{\Gamma}=\frac{p}{d^2}{\bf 1} +(1-p)P_+\ge 0
\end{equation}
$\widetilde E_{T}$ becomes separable at the point it becomes a
state. This implies that the structural approximation  to
transposition is always entanglement breaking.

Employing the $UU$-invariance of Werner states, we find an explicit expression for $\tilde T$
in the Holevo form \eqref{holevo}. Recall that each Werner state
can be represented using the $UU$ depolarizing map $\dep_{UU}$ as \cite{Werner}:
\begin{equation}\label{UU}
\varrho_W=\dep_{UU}(\varrho)=\int \de U (U\otimes U) \varrho
(U^{\dag}\otimes U^{\dag}) ,
\end{equation}
where $\de U$ corresponds to the Haar measure over the unitary group.
Since Werner states are spanned by the operators $\{\1,\V\}$
\cite{Werner}, normalized Werner states are completely defined by
the parameter $\ev{\V}=\tr(\varrho_W \V)$.

For the critical witness, i.e. $\widetilde E_{T}$ with minimal
$p$, we have $\ev{\V}=1$. One can easily check that the state
$\varrho=\proj{00}$ has the same expectation value, hence
\begin{equation}
\widetilde E_T=\int \de U \proj{v_U}\otimes\proj{w_U}
\end{equation}
with $\ket{v_U}=\ket{w_U}=U\ket{0}$. Notice that this expression
is a continuous version of \eqref{sep dec}, where the discrete set
of states $\{\ket{v_k},\ket{w_k}\}$ is replaced by a continuous
set $\{\ket{v_U},\ket{w_U}\}$ and the probability $p_k$ is
replaced by the probability distribution $\de U$. According to Eq.
\eqref{holevo_expl}, $\tilde T$ can be written as
\begin{equation}
\widetilde T(\varrho)= \int \de U
\proj{\bar w_U}\tr\Big[\left(d\proj{v_U}\right)\varrho\Big] ,
\end{equation}
where we used the invariance of the integral under conjugation. This approximation has a clear intuitive explanation. Given an unknown state, first one tries to estimate it in an optimal way using the covariance measurement defined by the infinite set of
operators $\{M_U=d\proj{v_U}\}$, distributed according to the Haar measure. If the measurement outcome corresponding to $\ket{v_U}$ is obtained, the state $\proj{v_U}^T=\proj{\bar w_U}$ is prepared.
Finally, it is important to mention that the map defining the depolarization process $\dep_{UU}$ can also be
implemented by the finite set of unitary operators $\{p_k,U_k\}$
of \cite{DCLB}, which in our case leads to a measurement with a
finite number of outcomes.

\subsection{Reduction Criterion}\label{ssec. reduction}
Finally, we consider the (normalized) reduction map $\Lambda_R$
defined as follows:
\begin{equation}\label{red}
\Lambda_R(\rho)=\frac{1}{d-1}\Big[\tr (\rho){\bf 1}-\rho\Big].
\end{equation}
which is also an optimal decomposable map \cite{reduction}. The
condition for the structural approximation $\widetilde\Lambda_R$ to
be completely positive reads:
\begin{equation}
{\bf 1}\otimes\widetilde \Lambda_R(P_+)\equiv\widetilde E_R=
\frac{d-p}{d^2(d-1)} {\bf 1}-\frac{1-p}{d-1}P_+\ge 0,
\end{equation}
which is immediately equivalent to:
\begin{equation}\label{spaR}
p\ge \frac{d}{d+1}.
\end{equation}
In order to study the separability of $\widetilde E_R$, note that
$\widetilde E_R$ is an isotropic state, i.e. $\widetilde E_R$ is
$U\bar U$-invariant. For such states the PPT criterion is again both
a necessary and sufficient condition for separability
\cite{reduction,werner_sym}. Denote by $\Pi_\pm$ the projectors onto
the symmetric $\text{Sym}(\mh\otimes\mh)$ and skew-symmetric
$\mh\wedge\mh$ subspaces respectively. With the help of the
identities $P_+=(1/d)\V^{\Gamma}$ and $\V=\Pi_+-\Pi_-$ one obtains
that:
\begin{equation}
\widetilde E_R^{\Gamma}=\frac{p}{d^2}\Pi_++
\frac{2d-p(d+1)}{d^2(d-1)}\Pi_-,
\end{equation}
which is positive for all $p$. Hence, when
$\widetilde E_R$ becomes positive, it also becomes separable which
implies that the structural approximation to $\Lambda_R$ is always
entanglement breaking.

Again we use the invariance properties of $\widetilde E_R$ to
write $\widetilde \Lambda_R$ in the Holevo form \eqref{holevo}.
These states belong to a space generated by $\{\1,P_+\}$ and
therefore can be completely described through the parameter
$\ev{P_+}=\tr(\varrho P_+)$. For the critical witness, the
expected value is $\ev{P_+}=0$ and a possible separable
decomposition of the state reads:
\begin{equation}
\widetilde E_R=\int \de U (U\otimes \bar U) \proj{\phi} (U\otimes
\bar U)^{\dag}
\end{equation}
with $\ket{\phi}=\ket{01}$. According to Eq.~\eqref{holevo_expl},
\begin{equation}
\widetilde \Lambda_R(\varrho)=\int dU \proj{w_{U}}\tr\left(\left(d\proj{v_U}\right)\varrho\right)
\end{equation}
where $\ket{v_{U}}=U\ket 0$ and $\ket{w_{U}}=U\ket1$.

\section{Non-decomposable maps}\label{nondec}

In this section, we move to non-decomposable maps. We first consider
the Choi map, which is one of the first examples of a
non-decomposable positive map. After this, we study those maps
coming from unextendible product bases. Finally, we analyze a
recently introduced positive map, the Breuer-Hall map. In all the
cases, we are able to prove the conjecture in arbitrary dimension.

\subsection{Choi's map} \label{sec. Choi} We now move to a
non-decomposable map proposed by Choi \cite{choi}. The normalized
map $\Lambda_C:\mathcal B(\mathbb C^3)\rightarrow \mathcal B
(\mathbb C^3)$ can be written as:
\begin{equation}\label{choi}
\Lambda_C(\rho)=\frac{1}{2}\left(-\rho+\sum_{i=0}^2
\rho_{ii}\left(2\proj{i}+\proj{i-1}\right)\right),
\end{equation}
where $\ket{i}$ is a fixed basis of $\mathbb C^3$ and the
summation is modulo 3. According to Eq.~\eqref{genform}, the
witness $\widetilde E_{C}$ associated with the structural
approximation $\widetilde \Lambda_C$ reads:
{\setlength\arraycolsep{2pt}
\begin{eqnarray}
&\widetilde E_C&= p\frac{\1}{9}+\\ \nonumber
&+&\frac{1-p}{6}\left(\sum_{i=0}^2\left[
2\proj{ii}+\proj{i,i-1}\right]-3P_+\right).
\end{eqnarray}}
By checking the positivity of this state we find that the map
$\widetilde \Lambda_C$ is completely positive for $p \geq 3/5$.

The entanglement witness $\widetilde E_C$ is separable since, for
critical $p$, it can be represented by the following convex
combination of (unnormalized) product states:
\begin{equation}\label{choiEcrit}
\widetilde
E_C=\frac{1}{15}\left(\sigma_{01}+\sigma_{12}+\sigma_{02}+
\sigma_d\right).
\end{equation}
Here $\sigma_d=\proj{02}+\proj{10}+\proj{21}$ is obviously
separable and the matrices $\sigma_{ij}$ are defined on the
subspace $ij$, i.e. spanned by
$\{\ket{ii},\ket{ij},\ket{ji},\ket{jj}\}$, and read:
\begin{equation}
\sigma_{ij}=\1-\ketbra{ii}{jj}-\ketbra{jj}{ii}.
\end{equation}
We can easily check that these density operators are PPT and hence
separable. Choi's map is not proven to be optimal, and there are
even reasons to believe that it is not. Namely, if one looks at
product vectors at which the mean of $E_C$ vanishes, they have the
form $(1, \exp(i\phi_1), \exp(i\phi_2))\otimes (1, \exp(-i\phi_1),
\exp(-i\phi_2))$, and are orthogonal to the vector
$(1,0,0,0,1,0,0,0,1)$, so that they  do not span the whole Hilbert
space and do not fulfill the assumptions of the Corollary 2 from
Section 2. Still, as we have shown, the  structural physical
approximation for the Choi's map is entanglement breaking. Thus, if
this map is (is not) optimal, this supports (does not contradict)
our conjecture.

\subsection{UPB Map}\label{UPB} Let us now focus on unextendible
product basis \cite{terhal_upb}. Recall that an unextendible
product basis in an arbitrary space $\mh_A\otimes\mh_B\cong\mathbb
C^{d_A}\otimes \mathbb C^{d_B}$ consists of a set of $n<d_Ad_B$
orthogonal product states, $\{v_i=x_i\otimes y_i\}_{i=1}^n$, such
that there is no product state orthogonal to them. It is then
impossible to extend this set into a full product basis. Given an
unextendible product basis, one can associate a PPT bound
entangled state
\begin{equation}
\varrho_{\{v\}}=\frac{1}{d_Ad_B-n}
\Big({\bf 1}_{AB}-\sum_{i=1}^n|v_i\rangle\langle v_i|\Big).
\end{equation}
The state is trivially PPT and entangled as there is no product state orthogonal to the UPB.

A (normalized) witness detecting such states can be taken in the
form \cite{terhal_upb}
\begin{equation}
\label{upbwitn}
E_{\{v\}}=\frac{1}{n-\epsilon
d_Ad_B}\left(\sum_{i=1}^n|v_i\rangle\langle v_i|-\epsilon {\bf
1}_{AB}\right),
\end{equation}
where $\epsilon>0$. A map $\Lambda_{\{v\}}\colon \mathcal
B(\mh_A)\to\mathcal B(\mh_B)$ corresponding to $E_{\{v\}}$ can be
obtained through the Jamio\l kowski's isomorphism (cf. Eq. \eqref{LE}) and
is a non-decomposable map since the state $\varrho_{\{v\}}$ is PPT. Let us consider the structural approximation $\widetilde
\Lambda_{\{v\}}$ to $\Lambda_{\{v\}}$. The witness associated with
$\widetilde \Lambda_{\{v\}}$ reads:
\begin{eqnarray}
&\widetilde E_{\{v\}}&=\frac{p}{d_Ad_B}{\bf 1} +(1-p)E_{\{v\}}\\
\nonumber &=&\frac{1}{n-\epsilon d_Ad_B}\left(\frac{np-\epsilon d_A
d_B}{d_Ad_B}{\bf 1} +(1-p)\sum_{i=1}^n|v_i\rangle\langle
v_i|\right).
\end{eqnarray}
Since by definition all vectors $v_i$ are product, $\widetilde
E_{\{v\}}={\bf 1}_{A}\otimes\widetilde \Lambda_{\{v\}}(P_+)$ is
separable, once it becomes positive. Therefore, structural
approximations to positive maps (\ref{upbwitn}) arising from UPB's
are entanglement breaking.

Since $\widetilde E_{\{v\}}$ is already in a product state
form, the Holevo representation comes directly from Eq.
\eqref{holevo_expl} for each particular unextendible product basis
$\{\ket{v_i}\}$. As mentioned, this gives the explicit construction of the measurement and state
preparation protocol approximating the map.

\subsection{Breuer-Hall Map}\label{HBs}
In what remains, we study the Breuer-Hall map, recently introduced in~\cite{Breuer,Hall}. This positive map can be understood
as the generalization of the reduction criterion to
the non-decomposable case. As we show next, this map also satisfies the conjecture for any dimension. When proving these results, we are
naturally led to the analysis of a new two-parameter family of
invariant states, that we call unitary symplectic invariant
states.

For even dimension $d=2n\ge 4$, which from this moment on we
assume, the reduction map (\ref{red}) can be yet improved, leading
to the (normalized) Breuer-Hall map \cite{Breuer, Hall}:
\begin{equation}\label{BH}
\Lambda_{BH}(\rho)=\frac{1}{d-2}\Big[\tr(\rho){\bf
1}-\rho-U\rho^TU^\dagger\Big].
\end{equation}
Here $U$ is any skew-symmetric unitary operator, i.e. $U^\dagger
U={\bf 1}$ and $U^T=-U$. The resulting map is no longer decomposable
and is known to be optimal \cite{Breuer}. From the general formula
(\ref{genform}), the entanglement witness associated with the structural
approximation $\widetilde\Lambda_{BH}$ is given by:
\begin{eqnarray}\label{EBH} \nonumber
\widetilde E_{BH}&=&\frac{1}{d-2}\bigg[\frac{d-2p}{d^2}{\bf 1}
-(1-p)P_+\\  & &-\frac{1-p}{d}\big({\bf 1}\otimes U\big) \V
\big({\bf 1}\otimes U^\dagger\big)\bigg].
%\nonumber\\
%&=&\frac{1}{d-2}\bigg[\frac{d-2p}{d^2}{\bf 1}
%-(1-p)P_++\frac{1-p}{d}\big({\bf 1}\otimes U\big) \V \big({\bf 1}\otimes \bar U\big)\bigg],
\end{eqnarray}

Further analysis of $\widetilde E_{BH}$ will be again based on
symmetry considerations. First of all, we note that since $U$ is
non-degenerate ($|\text{det}U|=1$) and skew-symmetric, there
exists a basis, known as Darboux basis, in which $U$ takes the
canonical form:
\begin{equation}\label{J}
J=\bigoplus_{i=1}^n\left[\begin{array}{cc} 0 & 1\\
                                -1 & 0\end{array}\right].
\end{equation}
For convenience, we choose $U=J$. Now, let $S\in
Sp(2n,\mathbb C)\cap U(2n)$ be a unitary symplectic matrix, i.e. a
complex matrix satisfying:
\begin{eqnarray}
S^\dagger S={\bf 1}\label{U}
\end{eqnarray}
and
\begin{equation}
SJS^T=J.\label{S}
\end{equation}
Then:
\begin{eqnarray}\nonumber
& &S\otimes \bar S\, \widetilde E_{BH}\, S^\dagger\otimes S^T\\
\nonumber& &=\alpha{\bf 1}+\beta\, S\otimes \bar S
P_+S^\dagger\otimes S^T +\gamma\, \big(S\otimes \bar SJ\big) \V
\big(S^\dagger\otimes J^\dagger S^T\big)\\ \nonumber & &=\alpha{\bf
1}+\beta\, P_++\gamma\, \big({\bf 1}\otimes J\big)\big(S\otimes
S\big) \V \big(S^\dagger\otimes S^\dagger\big)\big({\bf 1}\otimes
J^\dagger\big)\\ & &=\alpha{\bf 1}+\beta\, P_++\gamma\, \big({\bf
1}\otimes J\big) \V \big({\bf 1}\otimes J^\dagger\big)=\widetilde
E_{BH},
\end{eqnarray}
where we introduced constants $\alpha,\beta,\gamma$ for simplicity
(cf. Eq. (\ref{EBH})). In the second step above we used the
property:
\begin{equation}\label{SJ}
\bar S J=J S
\end{equation}
(it follows from Eqs. (\ref{U}), (\ref{S}) and the fact that $\bar
J=J$), together with the $U\bar U$-invariance of $P_+$
\cite{reduction}. Then in the last step we used the
$UU$-invariance of $\V$ \cite{Werner}.

Thus, we have just proven that the witness $\widetilde E_{BH}$,
associated with the Breuer-Hall map, is invariant under
transformations of the form $S\otimes \bar S$, with $S$ unitary
symplectic. Equivalently, its partial transpose $\widetilde
E_{BH}^\Gamma$ is invariant under transformations of the form
$S\otimes S$. We will generally call such operators unitary
symplectic invariant, or more specifically $S\bar S$- and
$SS$-invariant respectively.

To our knowledge, these operators have not been studied
systematically as an independent family. They form a subfamily
of $SU(2)$-invariant states of Ref. \cite{Breuer2}
(see also Ref. \cite{Breuer} where a subfamily of $SS$-invariant states was
introduced and Appendix B), but since the number of parameters of the latter
family increases with the dimensionality, it is manageable only
for low dimensions.
Below we describe unitary symplectic invariant states in any even dimension
(see e.g. Ref. \cite{werner_sym} for a general theory of
states invariant under the action of a group $G$). The results are then applied to the investigation of the
entanglement breaking properties of the structural approximations to the Breuer-Hall map.

\subsubsection{Unitary Symplectic Invariant States}\label{US}

In the next lines, we characterize the family of Unitary
Symplectic Invariant states. For the sake of clarity, here we state the main results, the corresponding proofs are then presented in Appendix C.

First of all, one should identify the space of Hermitian
$SS$-invariant operators. As shown in Appendix C, the spaces of
$SS$-invariant and $S\bar S$-invariant operators are \cite{note2}:
\begin{eqnarray}
SS\text{-inv}&\equiv&\text{Span}\{\1,\V,P_+^J\}\label{SS},\\
S\bar S\text{-inv}&\equiv&\text{Span}\{\1,P_+,\V^J\}
,\label{SSbar}
\end{eqnarray}
where $A^J\equiv({\bf 1}\otimes J)A({\bf 1}\otimes J^\dagger)$.

For a later convenience we introduce two equivalent sets of
generators given by the following minimal projectors:
\begin{eqnarray}\label{genS}
& &\Pi_0=P_+^J,\label{p1}\\
& &\Pi_1=\frac{1}{2}(\1-\V)-P_+^J,\label{p2}\\
& &\Pi_2=\frac{1}{2}(\1+\V)\label{p3},
\end{eqnarray}
and
\begin{eqnarray}\label{genSbar}
& &\hat \Pi_0=\Pi_0^J=P_+,\label{p1b}\\
& &\hat \Pi_1=\Pi_1^J=\frac{1}{2}(\1-\V^J)-P_+\label{p2b},\\
& &\hat \Pi_2=\Pi_2^J =\frac{1}{2}(\1+\V^J)\label{p3b}.
\end{eqnarray}
Relations (\ref{rel}-\ref{rel2}) imply that both sets define a
projective resolution of the identity:
\begin{equation}
\Pi_\alpha \Pi_\beta=\delta_{\alpha\beta}\Pi_\beta
\quad\text{and}\quad \sum_\alpha \Pi_\alpha=\1,
\end{equation}
and analogously for $\hat \Pi_\alpha$. Moreover $[\Pi_\alpha,\hat
\Pi_\beta]=0$ \cite{commut}.
%Projectors (\ref{p1}-\ref{p3})
%correspond to the decomposition of a $2n\times 2n$-matrix $M$ into
%the symplectic trace $\tr(JM)J$, the skew-symmetric symplectic
%traceless part $(1/2)(M-M^T)-\tr(JM)J$, and the symmetric part
%$(1/2)(M+M^T)$, respectively.

Projectors $\Pi_\alpha$ and $\hat \Pi_\alpha$ form extreme points of
the convex set of positive unitary symplectic invariant operators.
This allows us to easily describe the convex sets $\Sigma$ and $\hat
\Sigma$ of $SS$- and $S\bar S$-invariant states respectively. The
normalization implies that each family of states is uniquely
determined by two parameters: $\tr(\varrho \V), \tr(\varrho P_+^J)$
for $SS$-invariant states and $\tr(\varrho \V^J),\tr(\varrho P_+)$
for $S\bar S$-invariant ones (compare with Refs.
\cite{werner_sym,darek}, where orthogonal invariant states were
characterized). The extreme points of $\Sigma$ and $\hat \Sigma$ are
given by the normalized projectors $\Pi_\alpha/\tr\Pi_\alpha$ and
$\hat\Pi_\alpha/\tr\hat\Pi_\alpha$ respectively. We stress that both
sets live in two different subspaces of the big space of all
Hermitian operators: $\Sigma\subset\text{Span}_{\mathbb
R}\{\1,\V,P_+^J\}$ and $\hat \Sigma\subset\text{Span}_{\mathbb
R}\{\1,P_+,\V^J\}$. Partial transposition $\Gamma$ brings one set
into the plane of the other and allows one to study PPT and
separability.

\begin{figure}
\includegraphics[height=0.45\textwidth]{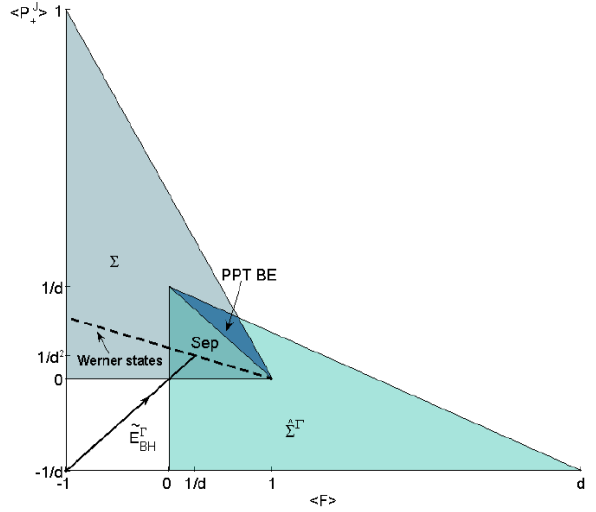}
\caption{The plot of the set $\Sigma$ of $SS$-invariant states
together with  $\hat \Sigma^\Gamma$---the set of
partial transposes of $S\bar S$-invariant states.
The thick line with the arrow represents
the partially transposed witness
$\widetilde E_{BH}^\Gamma(p)$. The dashed line represents
Werner states; its prolongation to the vertex
$(d,-1/d)\equiv\hat\Pi_0^\Gamma$ gives NPT
isotropic states. The family $\rho(\lambda)$
from Ref. \cite{Breuer} is given by the edge, connecting
vertices $(-1,1)\equiv\Pi_0$ and $(1,0)\equiv\Pi_2$.
The plot of $\hat\Sigma$, $\Sigma^\Gamma$, and
$\widetilde E_{BH}(p)$ is identical, with
the axes labels changed to $\langle\V^J\rangle$ and
$\langle P_+\rangle$ respectively.
} \label{figa1}
\end{figure}

Figure \ref{figa1} shows the plot of $\Sigma$
together with $\hat \Sigma^\Gamma$---the set of
partial transposes of $S\bar S$-invariant states.
For definiteness' sake
we have chosen to study partial transposes
of $S\bar S$-invariant states,
but as we will see the situation is fully symmetric.
The plane of the plot is the space of all Hermitian
$SS$-invariant operators with unit trace.
The set $\hat \Sigma^\Gamma$ is given by the convex hull
of the normalized operators
$\hat\Pi_\alpha^\Gamma/\tr\hat\Pi_\alpha$:
\begin{equation}
\hat \Sigma^\Gamma=\text{conv}\bigg\{\frac{\hat\Pi_1^\Gamma}{\tr\hat\Pi_1},
\frac{\hat\Pi_2^\Gamma}{\tr\hat\Pi_2},\frac{\hat\Pi_3^\Gamma}{\tr\hat\Pi_3}\bigg\}
\subset\text{Span}_{\mathbb R}\{\1,\V,P_+^J\}.\label{conv}
\end{equation}
The mentioned symmetry between the families
manifests itself in the fact that
by changing the axes labels $\langle\V\rangle\to\langle\V^J\rangle$
and $\langle P_+^J\rangle\to\langle P_+\rangle$ one obtains
the plot of $\hat \Sigma$ and $\Sigma^\Gamma$---
it is given by the identical figure in the
corresponding plane. This stems
from the following observations: $\tr(\hat\Pi_\alpha\V^J)=\tr(\Pi_\alpha\V)$,
$\tr(\hat\Pi_\alpha P_+)=\tr(\Pi_\alpha P_+^J)$,
$\tr\hat\Pi_\alpha=\tr\Pi_\alpha=\tr\hat\Pi_\alpha^\Gamma$, and
$\tr(\Pi_\alpha^\Gamma\V^J)=\tr(\hat\Pi_\alpha^\Gamma\V)$,
$\tr(\Pi_\alpha^\Gamma P_+)=\tr(\hat \Pi_\alpha^\Gamma P_+^J)$.

The intersection $\hat \Sigma^\Gamma\cap\Sigma$ describes those
$S\bar S$-invariant states with positive partial transpose. As
shown in Appendix C, not all of them are separable, i.e. there
are PPT entangled states in the family. The extreme points of the
intersection are given by:
\begin{eqnarray}\label{points}\nonumber
x_0&=&(0,0),\ x_1=\left(0,\frac{1}{d}\right),\\ x_2&=&(1,0),\
x_3=\left(\frac{d}{d+2},\frac{1}{d+2}\right).
\end{eqnarray}
To prove separability of a given point it is enough to show that
there exists a {\it normalized} product vector $\ket u\otimes \ket
v$ with the identical expectation values of $\V$ and $P_+^J$, for
the latter values characterize the state uniquely. Using this
fact, one can see that the extreme points of the separability
region are $x_0$, $x_1$ and $x_2$. The remaining part of the PPT
region contains entangled states.

\subsubsection{Entanglement breaking property of
$\widetilde\Lambda_{BH}$}

We can now return to the study of the witness $\widetilde E_{BH}$
associated with the Breuer-Hall map (cf. Eq. (\ref{EBH}) with
$U=J$). As we have shown in Section \ref{HBs}, $\widetilde E_{BH}$
is a $S\bar S$-invariant hermitean operator. Before analyzing when
it becomes positive, note that:
\begin{eqnarray}
\widetilde E_{BH}^{\Gamma}
&=&\frac{1}{d-2}\bigg[\frac{d-2p}{d^2}{\bf 1}
-\frac{1-p}{d}\V-(1-p)P_+^J\bigg]
\nonumber\\
&=&\big({\bf 1}\otimes J\big)\widetilde E_{BH}  \big({\bf 1}\otimes J^\dagger\big).
\label{EG=EJ}
\end{eqnarray}
Thus, $\widetilde E_{BH}\ge 0$ if and only if $\widetilde
E_{BH}^{\Gamma}\ge 0$, i.e. the structural-approximated witness is
a PPT state.

From Eqs.(\ref{EBH}) and (\ref{EG=EJ}) we obtain that when $0\le
p\le 1$:
\begin{eqnarray}
& &-1\le\tr(\widetilde E_{BH}\V^J)=\tr(\widetilde E_{BH}^\Gamma\V)
\le\frac{1}{d},\label{av1}\\
& &-\frac{1}{d}\le\tr(\widetilde E_{BH}P_+)=\tr(\widetilde E_{BH}^\Gamma P_+^J)
\le \frac{1}{d^2}.\label{av2}
\end{eqnarray}
The corresponding interval $p\mapsto\widetilde E_{BH}^\Gamma(p)$
is depicted in Fig.~\ref{figa1} by the thick line with the arrow.
We have plotted $\widetilde E_{BH}^\Gamma(p)$ rather than
$\widetilde E_{BH}(p)$. One sees that the line enters the positive
region $\Sigma$ at the point $x_0=(0,0)$, that is when both
averages (\ref{av1}) and (\ref{av2}) vanish. Equating any of the
expectation values to zero gives the condition for the structural
physical approximation:
\begin{equation}
p\ge \frac{d}{d+1}.
\end{equation}
Notice that it is the same bound as in Eq. (\ref{spaR}) for the
reduction map. Observing Fig.~\ref{figa1} it is clear that any
structural approximation to Breuer-Hall map is entanglement
breaking since the positivity region of $\widetilde E_{BH}^\Gamma$
is inside the separability region of SS-invariant states.

As a byproduct, we also obtain the minimum eigenvalue
$\lambda_{min}$ of the witness $E_{BH}$, corresponding to the
original positive map (\ref{BH}). From Eq.~(\ref{genform}) it
follows that at the critical probability $p=d/(d+1)$ one must have
$p/d^2+(1-p)\lambda_{min}=0$. This leads to $\lambda_{min}=-1/d$,
which corresponds to the eigenvector $\ket{\Phi_+}$. Note that
this eigenvector shares the symmetry of $E_{BH}$: $S\otimes\bar
S\ket{\Phi_+}=\ket{\Phi_+}$.

Again, we are able to provide a representation of the structural approximation to
Breuer-Hall map using the $S\bar S$-invariance of the corresponding
witness:
\begin{equation}
\widetilde E_{BH}=\int \de S (S\otimes \bar S)\proj{\varphi} (S\otimes
\bar S)^{\dag}.
\end{equation}
These states are parameterized by $\ev{P_+}$ and $\ev{\V^J}$ and,
for the critical witness $\widetilde E_{BH}$ we have
$\ev{P_+}=\ev{\V^J}=0$. The same expected values are obtained by
the separable state $\ket{\varphi}=\ket{\phi}\otimes \ket{\psi}$,
where
\begin{eqnarray}
\ket{\phi}&=&\frac{1}{2}\left(\ket{0}+\ket{1}+\ket{2}+\ket{3}\right)\\
\ket{\psi}&=&\frac{1}{\sqrt 2}\left(\ket{0}-\ket{2}\right).
\end{eqnarray}
Then, the Holevo form of $\widetilde \Lambda_{BH}$ is:
\begin{equation}
\widetilde \Lambda_{BH}(\varrho)=\int \de S \proj{w_{
S}}\tr\left(d\proj{v_S}\varrho\right)
\end{equation}
with $\ket{v_S}=S\ket{\phi}$ and $\ket{w_{S}}=S
\ket{\bar\psi}$.

\section{Entanglement Detection via Structural Approximations}
\label{entdet}

Before concluding, we would like to discuss the application of these ideas to the design of
entanglement detection methods. Indeed, one of the main motivations for the introduction
of structural approximations \cite{pawel_ekert} was to obtain approximate physical
realizations of positive maps, which can then be used for experimental entanglement detection.

\begin{figure}
\includegraphics[height=0.35\textwidth]{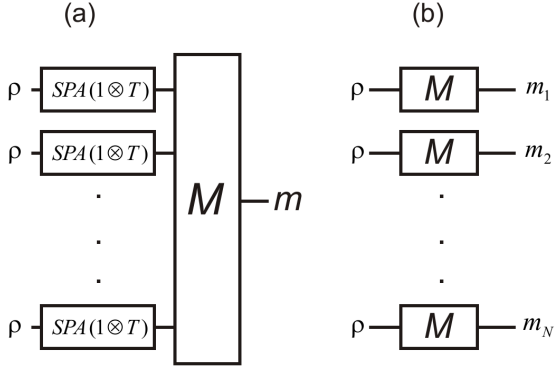}
\caption{The original scheme for direct entanglement detection
proposed in \cite{pawel_ekert} is shown in (a). Given $N$ copies of
an unknown state $\varrho$, it consists of, first, the structural
approximation of partial transposition acting on the initial state,
followed by optimal estimation of the minimal eigenvalue of the
resulting state. In the new scheme, all this structure is replaced
by single-copy measurements on the state. The minimal eigenvalue
should then be directly estimated from the obtained outcomes. }
\label{he_scheme}
\end{figure}

The original scheme proposed in \cite{pawel_ekert} works as follows,
see also Fig.~\ref{he_scheme}. Given $N$ copies of an unknown
bipartite state, $\varrho_{AB}$, the goal is to determine, without
resorting to full tomography, whether the state is PPT. The idea is
to apply the structural approximation to partial transposition to
this initial state and estimate the spectrum (or more precisely, the
minimal eigenvalue) of the resulting state using the optimal
measurement for spectrum estimation described in \cite{spectrest}.
Note that the structural approximation $\widetilde{\1\otimes T}$
``simply'' adds white noise to the ideal operator $\varrho^\Gamma$.
Thus, it is immediate to relate the spectrum of
$(\widetilde{\1\otimes T})(\varrho_{AB})$ to the positivity of the
partial transposition of the initial state.

Inspired by the previous findings, we study in this section
whether the structural approximation to partial transposition
defines an entanglement breaking channel. This map is of course
not even positive (so it does not entirely fit with our main
considered scenario), but obviously by adding sufficient amount of
noise it can be made not only positive but also completely
positive. As we show next, the structural approximation to partial
transposition does indeed define an entanglement breaking channel
whenever $d_A\geq d_B$, which includes the most relevant case of
equal dimension $d_A=d_B$.

This implies that the entanglement detection scheme of Fig.~\ref{he_scheme}.a
can just be replaced by a sequence of single-copy measurements, see Fig.~\ref{he_scheme}.b,
being the measurement the one associated to the Holevo form of the entanglement breaking channel. This
alternative scheme is much simpler from an implementation point of view since it
does not require any collective measurement, though the measurements are not projective.
Moreover, it can never be worse than the previous method, and most likely is better (see also \cite{remik}).

\subsection{Structural Approximations to $\1\otimes T$}

Let us then consider the structural
approximation to transposition extended to some arbitrary auxiliary
space: ${\bf 1}_A\otimes T_B$ \cite{pawel_ekert}. Note that, unlike
in the previous cases, the initial Hilbert space describing the
system is now explicitly a product
$\mh=\mh_A\otimes\mh_B\cong\mathbb C^{d_A}\otimes\mathbb C^{d_B}$.
Moreover, generally $\widetilde{{\bf 1}\otimes \Lambda}$ is not the
same as ${\bf 1}\otimes \widetilde \Lambda$, although
$\widetilde{{\bf 1}\otimes \Lambda}(P_+) ={\bf 1}\otimes \widetilde
\Lambda(P_+)$, so this problem does not reduce to the previous one.
Calculating the witness corresponding to $\widetilde{{\bf
1}_A\otimes T_B}$ one obtains:
\begin{eqnarray}
&\widetilde E_{{\bf 1}\otimes T}=\big[({\bf 1}_{A}\otimes{\bf
1}_{B})\otimes
(\widetilde{{\bf 1}_{A'}\otimes T_{B'}})\big](P_+^{AB,A'B'})\nonumber\\
&=\frac{p}{(d_Ad_B)^2} {\bf 1}_{AA'}\otimes{\bf
1}_{BB'}+\frac{1-p}{d_B}P_+^{AA'}\otimes \V^{BB'}, \label{E1T}
\end{eqnarray}
where $\V^{BB'}$ is the flip operator on $\mh_{B}\otimes
\mh_{B'}\cong \mathbb C^{d_B} \otimes \mathbb C^{d_B}$  and
$P_+^{AB,A'B'}$, $P_+^{AA'}$ are projectors onto maximally entangled
vectors in the corresponding spaces,
\begin{equation}
P_+^{AB,A'B'}=\frac{1}{d_Ad_B}\sum_{i,k=1}^{d_A}\sum_{j,l=1}^{d_B}
|ij\rangle_{AB}\langle kl|\otimes |ij\rangle_{A'B'}\langle kl|.
\end{equation}
The condition for structural approximation, positivity of
$\widetilde E_{{\bf 1}\otimes T}$, is most easily derived by using
the identity $\V^{BB'}=\Pi^{BB'}_+-\Pi^{BB'}_-$, where
$\Pi^{BB'}_+$ is the projector on the symmetric subspace
$\text{Sym}(\mh_{B}\otimes \mh_{B'})$, and introducing a projector
$Q_+^{AA'}={\bf 1}_{AA'}-P_+^{AA'}$. Then $\widetilde E_{{\bf
1}\otimes T}$ becomes: {\setlength\arraycolsep{2pt}
\begin{eqnarray}\nonumber
\widetilde E_{{\bf 1}\otimes T}&=&
\bigg[\frac{p}{(d_Ad_B)^2}+\frac{1-p}{d_B}\bigg]
P_+^{AA'}\otimes\Pi^{BB'}_+ \\
&+&\frac{p}{(d_Ad_B)^2}\Big[Q_+^{AA'}\otimes\Pi^{BB'}_+ +Q_+^{AA'}\otimes\Pi^{BB'}_-\Big]\nonumber\\
&+&\bigg[\frac{p}{(d_Ad_B)^2}-\frac{1-p}{d_B}\bigg]P_+^{AA'}\otimes\Pi^{BB'}_-.\label{pos}
\end{eqnarray}   }
Since only the last term can be negative, one obtains the following
condition for structural approximation:
\begin{equation}\label{1Tpos}
p\ge \frac{d_A^2d_B}{d_A^2d_B+1}.
\end{equation}
Comparison of the above threshold with the one given by Eq.
(\ref{cond}) with $d=d_B$, shows that in order to make ${\bf
1}_A\otimes T_B$ completely positive one has to add more noise than
to make the transposition $T$ alone completely positive and hence
implementable. In other words, ${\bf 1}_A\otimes \widetilde T_B$ is
less noisy than $\widetilde{{\bf 1}_A\otimes T_B}$.

We proceed to study the separability of $\widetilde E_{{\bf 1}\otimes T}$.
We begin by finding the partial transposition of $\widetilde E_{{\bf 1}\otimes T}$
with respect to the subsystem $A'B'$ \cite{note0}:
\begin{eqnarray}
\widetilde E_{{\bf 1}\otimes T}^{T_{A'B'}}=\frac{p}{(d_Ad_B)^2} {\bf
1}+\frac{1-p}{d_A}\V^{AA'}\otimes P_+^{BB'}.
\end{eqnarray}
Applying the same technique as above (cf. Eq. (\ref{pos})), we find that
$\widetilde E_{{\bf 1}\otimes T}^{\Gamma}\ge 0$ if and only if:
\begin{equation}
p\ge \frac{d_Ad_B^2}{d_Ad_B^2+1}.
\end{equation}
Comparing this to the threshold for positivity (\ref{1Tpos}),
we see that for $d_A < d_B$, i.e.  when the extension is
by a space of smaller dimension, there is
a gap between positivity and PPT. Hence, in this case,
for
\begin{equation}
 \frac{d_A^2 d_B}{d_A^2 d_B+1}\le p \le
\frac{d_Ad_B^2}{d_Ad_B^2+1}
\end{equation}
the witness (\ref{pos}) is not separable and the map
$\widetilde{{\bf 1}_A\otimes T_B}$ is not entanglement breaking in
this region. Recall however that this does not represent any counter-example
to the conjecture as the initial map is not even positive.

In the case $d_A \geq d_B$, we will use symmetry arguments to prove
the separability of $\widetilde E_{{\bf 1}\otimes T}$. From Eq.
(\ref{E1T}) it follows that this state is $U\bar UVV$-invariant, where
$U\in U(d_A)$, $V\in U(d_B)$
(cf. Refs. \cite{werner_sym, masanes} where $UUVV$-invariant states
were studied). Since both groups $U(d_A)$ and $U(d_B)$ act
independently it is easy to convince oneself \cite{werner_sym} that
the space of $U\bar UVV$-invariant operators is spanned by
$\{\1\otimes\1$, $\1\otimes \V$, $P_+\otimes \1$, $P_+\otimes \V$\}.
Following the same approach as in subsection \ref{ssec. trans}, we
prove the separability of $\widetilde E_{{\bf 1}\otimes T}$ in the
$AB:A'B'$ partition by showing that the state can be written as
convex sum of product states, i.e. it is has the following
representation
\begin{equation}
\int \de U \de V (U_A V_B\bar U_{A'} V_{B'})\sigma (U_A V_B \bar
U_{A'} V_{B'})^{\dag}
\end{equation}
(we omit tensor product signs here for brevity) for some $\sigma$
separable in the partition $AB:A'B'$. Given that states with this
invariance are completely described by parameters
$\ev{\1\otimes\V}$, $\ev{P_+\otimes\1}$ and $\ev{P_+\otimes\V}$,
$\sigma$ must obey the conditions: $\tr(\sigma
\1\otimes\V)=\tr(\widetilde E_{{\bf 1}\otimes T}\1\otimes\V)$,
$\tr(\sigma P_+\otimes\1)=\tr(\widetilde E_{{\bf 1}\otimes
T}P_+\otimes\1)$ and $\tr(\sigma P_+\otimes\V)=\tr(\widetilde
E_{{\bf 1}\otimes T}P_+\otimes\V)$. Such state
$\sigma\equiv\proj{\varphi}$ can be written as
\begin{eqnarray}\nonumber
\ket{\varphi}&\equiv&\ket{\phi}_{AB}\otimes
\ket{\psi}_{A'B'}\\
&=&\left(\sqrt{\alpha_{00}}\ket{00}+\sqrt{\alpha_{01}}\ket{01}+\sqrt{\alpha_{11}}\ket{11}\right)\ket{00}
\end{eqnarray}
for \begin{eqnarray}
\alpha_{00}&=&\frac{d_B}{d_Ad_B^2+1}(1+d_A)\\
\alpha_{01}&=&\frac{1}{d_Ad_B^2+1}( d_B^2+d_A-d_B(1+d_A)) \\
\alpha_{11}&=&1-\frac{1}{d_Ad_B^2+1}(d_B^2+d_A). \end{eqnarray}
Notice that, as expected, $\sigma$ is only well-defined for $d_A\geq
d_B$. According to Eq. \eqref{holevo_expl}, the map $\widetilde
{\1\otimes T}(\varrho)$ can be written as
\begin{equation} \label{1xT holevo}
\widetilde {\1\otimes T}(\varrho)=\int \de U\de V \proj{w_{U\bar V}}
\tr(d_A d_B\proj{v_{UV}}\varrho).
\end{equation}
where $\ket{v_{UV}}=U\otimes V\ket{\phi}$ and $\ket{w_{U\bar V}}=U\otimes\bar  V\ket{\bar \psi}$. Recall also that the integrals
over the unitary group defining each depolarization protocol can be
replaced by the finite sums of, e.g., Ref.~\cite{DCLB}.

In the case $d_A=d_B\equiv d$, we encounter the structural
approximation to the transposition map analyzed in
\cite{pawel_ekert}.
%The corresponding critical witness reads
%\begin{equation}
%\widetilde E_{{\bf 1}\otimes T} =\frac{1}{d(d^3+1)} \Big[{\bf
%1}_{AA'}\otimes{\bf 1}_{BB'}+P_+^{AA'}\otimes \V^{BB'}\Big].
%\end{equation}
%and has the following decomposition:
%\begin{equation}
%\widetilde E_{{\bf 1}\otimes T} = \int \de U \de V (U_A V_B \bar
%U_{A'}V_{B'})\proj{\varphi} (U_A V_B\bar U_{A'} V_{B'})^{\dag}
%\end{equation}
%where
%\begin{eqnarray}\nonumber
%\ket{\varphi}&\equiv&\ket{\phi}\otimes
%\ket{\psi}\\
%&=&\left(\alpha\ket{00}+\sqrt{1-\alpha}\ket{11}\right)_{AB}\ket{00}_{A'B'}
%\end{eqnarray}
%for $\alpha=d(d+1)/(d^3+1)$.
As mentioned, by providing the representation \eqref{1xT holevo} we
are able to replace the former entanglement detection scheme
\cite{pawel_ekert} by a much less resource-demanding one. In the
original proposal, $n$ copies of $\tilde T(\varrho)$ are prepared,
followed by optimal estimation of its minimal eigenvalue by means of a collective projective
measurement on the $n$-copy state. Now, one should just perform local
measurements in the $n$ copies of $\varrho$ with operators
defined in \eqref{1xT holevo} and with that directly estimate the
lowest eigenvalue of $\1\otimes T$.

\section{Conclusions}\label{concl}

In this work, we have studied the implementation of structural
approximations to positive maps via measurement and
state-preparation protocols. Our findings suggest an intriguing
connection between these two concepts that we have summarized by
conjecturing that the structural physical approximation of an
optimal positive map defines an entanglement breaking channel. Of
course, the main open question is (dis)proving this conjecture. It
would also be interesting to obtain slightly weaker results in the
same direction, such as proving the conjecture for general optimal
decomposable maps (which seems more plausible due to the fact that
the conjecture holds for transposition). We have also applied the
same ideas to the study of physical approximations to partial
transposition, which is not a positive map, and discuss the
implications of our results for entanglement detection.

\begin{figure}
 \includegraphics[height=0.35\textwidth]{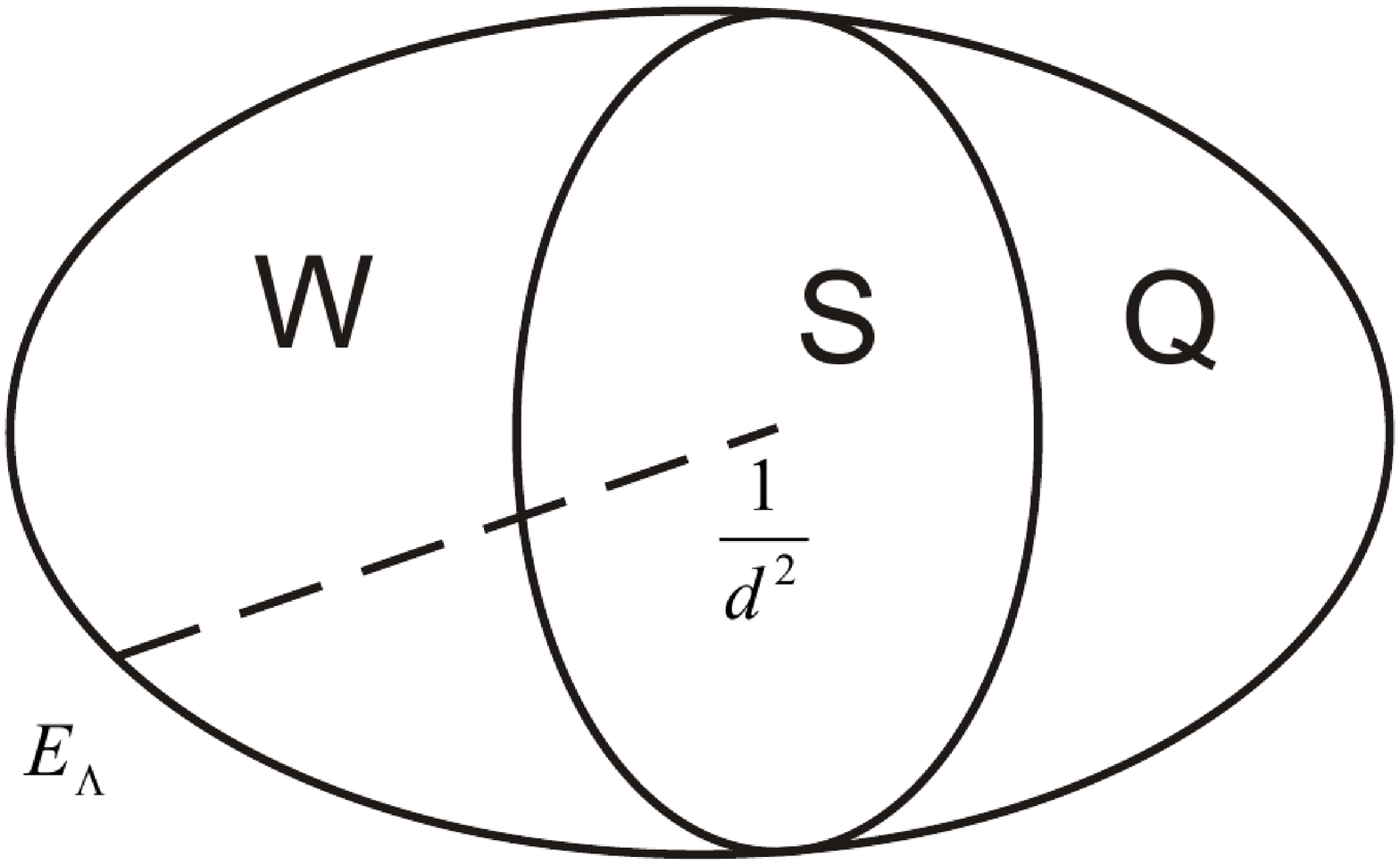}
  \caption{The sets $S$, $Q$ and $W$ of separable states, quantum states and operators positive
on product states are such that $S\subset Q\subset W$. If the conjecture was true, namely all
structural approximations to optimal positive maps defined entanglement breaking channels, it
would mean that optimal positive maps (witnesses) enter the physical region, when adding white noise,
via the separability region, as shown in the figure.}\label{geom}
\end{figure}

We would like to conclude this work by giving a geometrical
representation of our findings (that should be interpreted in an
approximate way). It is well known that the set of quantum states
is convex and includes the set of separable states, which is also
convex, see also Fig.~\ref{geom}. These two sets are contained in
the set of Hermitian operators that are positive on product
states, which is again convex. Entanglement witnesses belong to
this set. If the conjecture was true, it would mean that the set
of optimal witnesses would live in a region which is ``opposite"
to the set of separable states, in the sense that when mixed with
the maximally mixed noise, they enter the set of physical states
via the separability region.

Finally, let us mention some further open questions. It would be
interesting to extend our studies and ask which classes of positive
maps have structural approximation that corresponding to partially
breaking channels (for definition see \cite{darek1})? Is our
conjecture true for maps that are not optimal, but atomic
\cite{darek2}, i.e. detect Schmidt number 2 entanglement (for
definition see \cite{SBL}? What is the relation between optimality,
extremality (in the sense of convex sets) and atomic property?

\acknowledgements We would like to thank M., P., and R. Horodecki,
and A. Kossakowski, D. Chru\'sci\'nski, and S. Iblisdir for
discussions. We gratefully acknowledge the financial support from EU
Programmes ``SCALA'' and QAP, ESF PESC Programme ``QUDEDIS, Spanish
MEC grants (FIS 2005-04627, FIS2007-60182 and Consolider Ingenio
2010 ``QOIT), the Grup Consolidat de Recerca de la Generalitat de
Catalunya, Caixa Manresa, the Funda\c{c}\~{a}o para a Ci\^{e}ncia e
a Tecnologia (Portugal) through the grant SFRH/BD/21915/2005, and
the IT R$\&$D program of MKE/IITA (2008-F-035-01).

\section*{Appendix A: Proof of the conjecture for a rank-three optimal witness in $2\otimes 4$ systems}

In this appendix, we show that the structural approximation to the
optimal witness $Q^\Gamma$, where $Q$ is the projector onto states
\eqref{2x4vectors}, is separable. Following our general procedure
(cf. Eq. (\ref{genform})), the normalized witness associated to
the structural approximation reads:
\begin{eqnarray}
& &\widetilde E_\Lambda=\frac{p}{8}{\bf 1}+
\frac{1-p}{6}Q^\Gamma\equiv\frac{1-p}{6}\Big(Q^\Gamma+a{\bf 1}\Big)\nonumber\\
& &=\frac{1-p}{6}\times\nonumber\\
& &\left[\begin{array}{cccccccc} a & 0 & 0 & 0 & 0 & -1 & 0 & 0\\
                              0 & 1+a & 0 & 0 & 0 & 0 & -1 & 0\\
                              0 & 0 & 1+a & 0 & 0 & 0 & 0 &-1\\
                              0 & 0 & 0 & 1+a & 0 & 0 & 0 & 0\\
                              0 & 0 & 0 & 0 & 1+a & 0 & 0 & 0\\
                              -1 & 0 & 0 & 0 & 0 & 1+a & 0 & 0\\
                              0 & -1 & 0 & 0 & 0 & 0 & 1+a & 0\\
                              0 & 0 & -1 & 0 & 0 & 0 & 0 & a\end{array}\right]
\label{kurwiozum},
\end{eqnarray}
where $a=\frac{6p}{8(1-p)}$.
The above operator becomes positive when
\begin{equation}\label{posa}
a(1+a)=1.
\end{equation}
%which translates through
%Eq. (\ref{a}) into critical
%probability
%\begin{equation}
%p_c=\frac{8-2\sqrt 2}{7}\approx 0.73.
%\end{equation}
To show that at this point the matrix
(\ref{kurwiozum}) becomes separable, we first
perform a local invertible transformation
and pass from $Q^\Gamma+a{\bf 1}$ to
${\bf 1}\otimes A\Big(Q^\Gamma+a{\bf 1}\Big)
{\bf 1}\otimes A^\dagger$, where
\begin{equation}
A=\left[\begin{array}{cccc} 1 & 0 & 0 & 0\\
                           0 & a & 0 & 0\\
                           0 & 0 & a & 0\\
                           0 & 0 & 0 & 1 \end{array}\right].
\end{equation}
With the help of the positivity condition
(\ref{posa}),
the resulting matrix can be written as:
\begin{eqnarray}
& &a^2\left[\begin{array}{cccccccc} 1 & 0 & 0 & 0 & 0 & -1 & 0 & 0\\
                              0 & 1 & 0 & 0 & 0 & 0 & -1 & 0\\
                              0 & 0 & 1 & 0 & 0 & 0 & 0 & -1\\
                              0 & 0 & 0 & 1 & 0 & 0 & 0 & 0\\
                              0 & 0 & 0 & 0 & 1 & 0 & 0 & 0\\
                              -1 & 0 & 0 & 0 & 0 & 1 & 0 & 0\\
                              0 & -1 & 0 & 0 & 0 & 0 & 1 & 0\\
                              0 & 0 & -1 & 0 & 0 & 0 & 0 & 1\end{array}\right]+
\nonumber\\
& &+(a-a^2)
\left[\begin{array}{cccccccc} 1 & 0 & 0 & 0 & 0 & -1 & 0 & 0\\
                              0 & 1 & 0 & 0 & 0 & 0 & 0 & 0\\
                              0 & 0 & 1 & 0 & 0 & 0 & 0 & -1\\
                              0 & 0 & 0 & \kappa & 0 & 0 & 0 & 0\\
                              0 & 0 & 0 & 0 & \kappa & 0 & 0 & 0\\
                              -1 & 0 & 0 & 0 & 0 & 1 & 0 & 0\\
                              0 & 0 & 0 & 0 & 0 & 0 & 1 & 0\\
                              0 & 0 & -1 & 0 & 0 & 0 & 0 & 1\end{array}\right],
\label{last}
\end{eqnarray}
where
\be
\kappa=\frac{1}{a-a^2}\Big(\frac{1}{a}-a^2\Big)>1.
\ee
Note that since at the critical point (\ref{posa}), $a-a^2>0$,
it is enough to show that both matrices in the
above decomposition are separable.
The first matrix, which we denote by $\sigma$,
possesses the following continuous
separable representation:
\begin{equation}
\sigma=\int_0^{2\pi}\frac{\text{d}\phi}{2\pi}|\psi(\phi)\rangle\langle\psi(\phi)|,
\end{equation}
where
\begin{equation}
\psi(\phi)=(\text{e}^{\text{i}\phi}, -1)\otimes
(1, \text{e}^{\text{i}\phi},\text{e}^{2\text{i}\phi},\text{e}^{3\text{i}\phi}).
\end{equation}
The second matrix has a $(2\otimes 2)\oplus(2\otimes 2)$
structure with $2\otimes 2$ blocks being identical and given by
\be
\left[\begin{array}{cccc} 1 & 0 & 0 & -1\\
                           0 & 1 & 0 & 0\\
                           0 & 0 & \kappa & 0\\
                           -1 & 0 & 0 & 1 \end{array}\right].
\ee
Since $\kappa>1$ the above matrix is PPT and hence separable.
Thus, the whole matrix (\ref{last}) is separable, which finishes
the proof.

\section*{Appendix B: $3\otimes 3$ systems}\label{app33}

In this appendix, we provide several examples of positive maps
satisfying the conjecture. Again we consider decomposable optimal
maps and study case-by-case various possible ranks of the $Q$
operator (cf. Theorem 2, Section \ref{DW}).

The case $r(Q)=1$, i.e. $Q=|\psi\rangle\langle\psi|$, splits into
two subcases. When the Schmidt-rank of $\ket{\psi}$ is 2, $Q$ is
supported in a $2\otimes 2$ subspace and the structural
approximation is entanglement breaking by the previous results (cf.
Section \ref{2x2}). In the case where $\ket{\psi}$ is Schmidt-rank
3, we restrict our attention to the trace-preserving case, i.e.
assume that $\ket{\psi}$ is maximally entangled. Alternatively,
before checking the conjecture we apply local transformations and
bring $\ket{\psi}$ to the form (\ref{P+}), i.e. we assume that:
\begin{equation}
\ket{\psi}=\frac{1}{\sqrt 3}\Big(|00\rangle + |11\rangle +
|22\rangle \Big)=\ket{\Phi_+}.
\end{equation}
Then the corresponding witness $\tilde E$ from Eq. (\ref{genform})
turns out to be a Werner state \cite{Werner} of dimension $d=3$.
This witness was already studied for arbitrary $d$ in section
\label{ssec. trans}, where we concluded that such structural
approximation is always entanglement breaking.

%\begin{equation}
%\widetilde E_\Lambda=\frac{p}{9}{\bf 1}+(1-p)P_+^\Gamma
%=\frac{p}{9}{\bf 1}+\frac{1-p}{3}\V.
%\end{equation}
%Here $\V \psi\otimes\phi=\phi\otimes\psi$ is the flip operator and
%we used the fact that $\V^\Gamma=dP_+$. Obviously $\widetilde
%E_\Lambda^\Gamma\ge 0$ for all $p$ and since PPT criterion is both
%necessary and sufficient for separability of Werner states
%\cite{Werner}, $\widetilde E_\Lambda$ becomes separable at the
%point it becoms positive. The latter happens for $p\ge 3/4$ which
%is the condition for the structural approximation in this case.

We move to the case $r(Q)=2$. Then $Q$ has to be supported either
in a $2\otimes 3$ subspace or in the full $3\otimes 3$ space,
since in $2\otimes 2$ there is always a product vector in every
two dimensional subspace and $Q$ would not be optimal by Theorem 2
of Section \ref {DW}. The first case, when $Q$ is supported in a
$2\otimes 3$ subspace, is covered by Section \ref{2x2}. In the
other case, we do not have  a general theory, but in a generic
case the range of $Q$ is spanned by two Schmidt-rank 2 vectors. We
can take them to be:
\begin{eqnarray}
& &|01\rangle-|10\rangle\label{v2.1}\\
& &|12\rangle-|21\rangle\label{v2.2}.
\end{eqnarray}
Obviously, for such a $Q$ it holds $Qe\otimes e=0\Rightarrow
\langle e\otimes\bar e|Q^\Gamma e\otimes\bar e\rangle=0$ for any
$e\in\mathbb C^3$. Since vectors $e\otimes\bar e$ span the whole
$\mathbb C^3\otimes \mathbb C^3$, by Corollary 2 of Section
\ref{GO} the witness $Q^\Gamma$ is optimal.

Again we do not have a general result here, but only consider a
generic example of $Q$ given by the projectors on the above
vectors (\ref{v2.1}-\ref{v2.2}). The normalized witness
corresponding to the structural approximation, $\widetilde
E_\Lambda=\frac{1-p}{4}\Big(Q^\Gamma+a{\bf 1}\Big)$ with
$a=\frac{4p}{9(1-p)}$ is given, modulo the $(1-p)/4$ prefactor, by
the matrix:
\begin{equation}\label{swolocz}
\left[\begin{array}{ccccccccc} a & 0 & 0 & 0 & -1 & 0 & 0 & 0 & 0\\
                              0 & 1+a & 0 & 0 & 0 & 0 & 0 & 0 & 0\\
                              0 & 0 & a & 0 & 0 & 0 & 0 & 0 & 0\\
                              0 & 0 & 0 & 1+a & 0 & 0 & 0 & 0 & 0\\
                              -1 & 0 & 0 & 0 & a & 0 & 0 & 0 & -1\\
                              0 & 0 & 0 & 0 & 0 & 1+a & 0 & 0 & 0\\
                              0 & 0 & 0 & 0 & 0 & 0 & a & 0 & 0\\
                              0 & 0 & 0 & 0 & 0 & 0 & 0 & 1+a & 0\\
                              0 & 0 & 0 & 0 & -1 & 0 & 0 & 0 & a\end{array}\right].
\end{equation}
It becomes positive at the point $a(a^2-2)=0$, i.e. at
\begin{equation}\label{pizda}
a=\sqrt 2.
\end{equation}
which gives the critical probability $ p_c=\frac{9\sqrt 2}{9\sqrt
2+1}\approx 0.93.$

To check the separability at the above point (\ref{pizda}), note
that the matrix (\ref{swolocz}) can be decomposed as follows:
\begin{eqnarray}
& &\left[\begin{array}{ccccccccc} a & 0 & 0 & 0 & -1 & 0 & 0 & 0 & 0\\
                              0 & 1+a & 0 & 0 & 0 & 0 & 0 & 0 & 0\\
                              0 & 0 & 0 & 0 & 0 & 0 & 0 & 0 & 0\\
                              0 & 0 & 0 & 1+a & 0 & 0 & 0 & 0 & 0\\
                              -1 & 0 & 0 & 0 & \frac{a}{2} & 0 & 0 & 0 & 0\\
                              0 & 0 & 0 & 0 & 0 & 0 & 0 & 0 & 0\\
                              0 & 0 & 0 & 0 & 0 & 0 & 0 & 0 & 0\\
                              0 & 0 & 0 & 0 & 0 & 0 & 0 & 0 & 0\\
                              0 & 0 & 0 & 0 & 0 & 0 & 0 & 0 & 0\end{array}\right]+
\nonumber\\
& &+\left[\begin{array}{ccccccccc} 0 & 0 & 0 & 0 & 0 & 0 & 0 & 0 & 0\\
                              0 & 0 & 0 & 0 & 0 & 0 & 0 & 0 & 0\\
                              0 & 0 & 0 & 0 & 0 & 0 & 0 & 0 & 0\\
                              0 & 0 & 0 & 0 & 0 & 0 & 0 & 0 & 0\\
                              0 & 0 & 0 & 0 & \frac{a}{2} & 0 & 0 & 0 & -1\\
                              0 & 0 & 0 & 0 & 0 & 1+a & 0 & 0 & 0\\
                              0 & 0 & 0 & 0 & 0 & 0 & 0 & 0 & 0\\
                              0 & 0 & 0 & 0 & 0 & 0 & 0 & 1+a & 0\\
                              0 & 0 & 0 & 0 & -1 & 0 & 0 & 0 & a\end{array}\right]
+\nonumber\\
& &+\left[\begin{array}{ccccccccc} 0 & 0 & 0 & 0 & 0 & 0 & 0 & 0 & 0\\
                              0 & 0 & 0 & 0 & 0 & 0 & 0 & 0 & 0\\
                              0 & 0 & a & 0 & 0 & 0 & 0 & 0 & 0\\
                              0 & 0 & 0 & 0 & 0 & 0 & 0 & 0 & 0\\
                              0 & 0 & 0 & 0 & 0 & 0 & 0 & 0 & 0\\
                              0 & 0 & 0 & 0 & 0 & 0 & 0 & 0 & 0\\
                              0 & 0 & 0 & 0 & 0 & 0 & a & 0 & 0\\
                              0 & 0 & 0 & 0 & 0 & 0 & 0 & 0 & 0\\
                              0 & 0 & 0 & 0 & 0 & 0 & 0 & 0 & 0\end{array}\right].
\end{eqnarray}
The first two matrices are supported in $2\otimes 2$ subspaces.
Their partial transposes become positive for $(1+a)^2=1$, which is
satisfied at the point (\ref{pizda}). The last matrix is obviously
separable. This allows us to conclude that the structural
approximation (\ref{swolocz}) is entanglement breaking.

Next, we consider the case $r(Q)=3$. Then $Q$ must be supported in
the whole $3\otimes 3$ space (otherwise there would be a product
vector in the range of $Q$ and $Q$ would not be optimal by Theorem
2, Section \ref{DW}). In lieu of a general theory, we consider a
seemingly generic example of
\begin{equation}
Q=\Pi_-,
\end{equation}
where by $\Pi_\pm$ we denote the projectors onto the symmetric
$\text{Sym}(\mh\otimes\mh)$ and skew-symmetric $\mh\wedge\mh$
subspaces respectively. The corresponding normalized witness reads:
\begin{equation}
\frac{3}{1-p}\widetilde E_\Lambda=Q^\Gamma+a{\bf 1}
=\frac{1}{2}\Big[(1+2a){\bf 1}-3P_+\Big],\label{Q3}
\end{equation}
where $a=\frac{3p}{9(1-p)}$ and we used the identities
$\V=\Pi_+-\Pi_-=1-2\Pi_-$ and $\V^\Gamma=dP_+$. The condition for
structural approximation, $\widetilde E_\Lambda\ge 0$, is
equivalent to \be a\ge 1. \ee
%which gives
%the critical probability
%\begin{equation}
%p_c=\frac{9}{10}.
%\end{equation}
Note that the structural-approximated witness (\ref{Q3}) is an
isotropic state of dimension $d=3$ and that this was already studied
for arbitrary $d$ in Sec.~\ref{ssec. reduction}. There we concluded
that such witnesses always correspond to entanglement-breaking
channels.

We are left with the last case $r(Q)=4$. Note that generically if we
consider $P$ a projector on the kernel of $Q$, then $r(P)=5$ and the
range of $P$ contains exactly $\le$ 5 product vectors. In general,
$Q$ will contain some product vector in its kernel and therefore is
not optimal.
%These 5 (or less) product vectors evidently do not span the Hilbert
%space, so $Q^\Gamma$ does not necessarily fulfill the assumptions of
%Corollary 2 of Section II, and most likely is not optimal.
For this reason, here we consider not a generic but a particular $Q$
where optimality is guaranteed by the Corollary 2 of Section
\ref{GO}. We can treat $\mathbb C^3\otimes\mathbb C^3$ as a
representation space of two spin-$1$ representations of $SU(2)$. We
then consider positive operators $Q$ supported on a span of the
skew-symmetric subspace $\mathbb C^3\wedge\mathbb C^3$ and the
singlet \cite{Breuer2}: \be \Psi=\frac{1}{\sqrt
3}\Big(|02\rangle+|20\rangle -|11\rangle\Big). \ee Denoting by $J$
the total spin, $Q$ is supported on the sum of $J=0$ and $J=1$
subspaces, while $P$ is supported on the $J=2$ subspace. The kernel
of $Q$ is then spanned by the vectors of the form $(1,\sqrt
2\alpha,\alpha^2)\otimes(1,\sqrt 2\alpha,\alpha^2)$ for a complex
$\alpha$. By Corollary 2 of Section \ref{GO}, $Q^\Gamma$ is optimal,
as vectors $(1,\sqrt 2\alpha,\alpha^2) \otimes(1,\sqrt
2\bar\alpha,\bar\alpha^2)$ span whole of the $\mathbb C^3\otimes
\mathbb C^3$.

As a particular example we consider \be
Q=2\Pi_-+2P_\Psi.\label{qrwa} \ee The structural approximation
gives:
\begin{widetext}
\begin{eqnarray}
& &\frac{8}{1-p}\widetilde E_\Lambda=Q^\Gamma+a{\bf 1}=\nonumber\\
& &\left[\begin{array}{ccccccccc} a & 0 & 0 & 0 & -1 & 0 & 0 & 0 & -\frac{1}{3}\\
                              0 & 1+a & 0 & 0 & 0 & -\frac{2}{3} & 0 & 0 & 0\\
                              0 & 0 & \frac{5}{3}+a & 0 & 0 & 0 & 0 & 0 & 0\\
                              0 & 0 & 0 & 1+a & 0 & 0 & 0 & -\frac{2}{3} & 0\\
                              -1 & 0 & 0 & 0 & \frac{2}{3}+a & 0 & 0 & 0 & -1\\
                              0 & -\frac{2}{3} & 0 & 0 & 0 & 1+a & 0 & 0 & 0\\
                              0 & 0 & 0 & 0 & 0 & 0 & \frac{5}{3}+a & 0 & 0\\
                              0 & 0 & 0 & -\frac{2}{3} & 0 & 0 & 0 & 1+a & 0\\
                              -\frac{1}{3} & 0 & 0 & 0 & -1 & 0 & 0 & 0 & a\end{array}\right],
\label{chujoza}
\end{eqnarray}
\end{widetext}
where \be a=\frac{8p}{9(1-p)}. \ee The matrix (\ref{chujoza})
becomes positive at the point given by the conditions
$(a+1)^2-\frac{4}{9}=0$ and
$\Big(a+\frac{2}{3}\Big)\Big(a-\frac{1}{3}\Big)-2=0$, which is
solved by \be a=\frac{4}{3}. \ee

We now prove that at this point the witness (\ref{chujoza}) becomes
separable. We consider the partially transposed witness: \be
\frac{8}{1-p}\widetilde E_\Lambda^\Gamma =\frac{4}{3}{\bf
1}+2\Big(P_{J=1}+P_{J=0}\Big),\label{EwithPs} \ee where $P_{J}$
projects on the subspace of total spin $J$. Using the technique
based on the state invariance described in Sec.~\ref{trans}, we
explicitly construct a separable decomposition for $\widetilde
E_\Lambda^\Gamma$. Analogously to the definition \eqref{UU}, we
introduce spin-$1\,\otimes\,$spin-$1$ depolarizing operator:
\begin{eqnarray}
& &\mathscr{D}(\varrho)=\int\de \mathscr D^{(1)}(U)\times\nonumber\\
& &\big[\mathscr D^{(1)}(U)\otimes \mathscr D^{(1)}(U)\big]
\varrho \big[\mathscr D^{(1)}(U)^\dagger\otimes
\mathscr D^{(1)}(U)^\dagger\big]\\
& &=\frac{1}{5}\tr\big(\varrho P_{J=2}\big)P_{J=2}
+\frac{1}{3}\tr\big(\varrho P_{J=1}\big)P_{J=2}+\nonumber\\
& &+\tr\big(\varrho P_{J=0}\big)P_{J=0}.
\end{eqnarray}
where $\mathscr D^{(1)}(U)\in SO(3)$ denotes spin-$1$
representation of $U\in SU(2)$. By direct calculation we check
that \be \mathscr{D}(\proj{02})+\mathscr{D}(\proj{01}) \ee gives,
up to a positive constant, the desired operator $\widetilde
E_\Lambda^\Gamma$. Since separability of $\widetilde
E_\Lambda^\Gamma$ is equivalent to separability of $\widetilde
E_\Lambda$, we have thus shown that the structural approximation
to the map defined by Eq. (\ref{qrwa}) is entanglement breaking.

\section*{Appendix C: Analysis of Unitary Symplectic Invariant States}

The scope of this appendix is to provide a characterization of the
properties of $SS$ and $S\bar S$ invariant states. The first step is
to find the space of Hermitian $SS$-invariant operators. The
corresponding space of $S\bar S$-invariant ones is related to the
latter by partial transposition $\Gamma$. Since unitary symplectic
transformations $S$ are obviously unitary, all $UU$-invariant
operators are also $SS$-invariant. As it is well known, the former
space is spanned by $\1$ and $\V$ \cite{Werner}. As a rule,
shrinking the group enlarges the space of the invariant operators,
so one expects more than that. The form of the invariance group
$G=Sp(2n,\mathbb C)\cap U(2n)$ implies that
$\{G-\text{inv}\}=\{Sp(2n,\mathbb
C)-\text{inv}\}\cup\{U(2n)-\text{inv}\}$ (in some sense we will not
specify here; see Ref. \cite{werner_sym}). Thus, one has to find the
$Sp(2n,\mathbb C)$-invariant operators.

Let $A$ be Hermitian and such that:
\begin{equation}\label{Sinv}
\sum_{j,\dots,n} S_{ij}S_{kl}A_{jlmn}\bar S_{rm}\bar S_{sn}=A_{ikrs},
\end{equation}
for all $S$ from $Sp(2n,\mathbb C)$ (now $S$ satisfies Eq. (\ref{S}) only).
Since $S$ and its complex conjugation
$\bar S$ are independent for a general $S\in Sp(2n,\mathbb C)$,
and the defining equation (\ref{S}) does not involve complex conjugation,
the only possibility for Eq. (\ref{Sinv}) to hold is when $A$ is rank one, i.e.
$A_{jlmn}=\psi_{jl}\bar\phi_{mn}$. Then Eq. (\ref{Sinv}) becomes:
\begin{equation}\label{Sinv2}
\big(S\psi S^T\big)_{ik}\overline{\big(S\phi S^T\big)}_{rs}=\psi_{ik}\bar\phi_{rs}.
\end{equation}
But the only quadratic form that $S$ preserves is
$J$, which implies that one must have
$\psi_{ik}=c_1J_{ik}$ and $\phi_{rs}=c_2J_{rs}$ for some complex $c_{1,2}\ne 0$.
We choose $c_1=c_2=-1/\sqrt d$, $d=2n$, which leads to:
\begin{eqnarray}\label{P+J} \nonumber
\ket{\psi}&=&\ket{\phi}=-\frac{1}{\sqrt d}\sum_{i,k}J_{ik}|ik\rangle\\ \nonumber
&=&\frac{1}{\sqrt
d}\big(|10\rangle-|01\rangle+|32\rangle-|23\rangle+\dots\big)\\
 &=&(\1\otimes J)\ket{\Phi_+} ,
\end{eqnarray}
(cf. Eq. (\ref{P+})). Hence, $P_+^J=({\bf 1}\otimes J)P_+({\bf
1}\otimes J^\dagger)$ is the only $Sp(2n,\mathbb C)$-invariant
operator, up to a multiplicative constant \cite{note1}. Using this
fact we conclude that the space of $SS$-invariant operators is
spanned by $\{\1,\V,P_+^J\}$. Correspondingly, the space of $S\bar
S$-invariant operators is spanned by
$\{\1,\V,P_+^J\}^\Gamma\equiv\{\1,P_+,\V^J\}= (\1\otimes
J)\{\1,P_+^J,\V^J\}(\1\otimes J^\dagger)$. As a side remark, we note
that since $J$ is real, $J^\dagger=J^T=-J$ (cf. definition
(\ref{J})) and hence $(\1\otimes J)A(\1\otimes
J^\dagger)=-(\1\otimes J)A(\1\otimes J)$ for any $A$. We will use
this fact frequently, but keep writing $J^\dagger$.

As a general rule, $G$-invariant operators form
algebras \cite{werner_sym}. The constituent relations for
the algebras of unitary symplectic invariant operators are as follows:
\begin{eqnarray}\label{rel}
\V P_+^J&=&-P_+^J=P_+^J\V\quad\\ \nonumber & &\text{and}\\
\label{rel2}\quad P_+\V^J&=&-P_+=\V^JP_+.
\end{eqnarray}
The above relations follow from the identity
$\V(\1\otimes J)\ket{\Phi_+}=(J\otimes \1)\ket{\Phi_+}
=-(\1\otimes J)\ket{\Phi_+}$, equivalent to $\V^J\ket{\Phi_+}=-\ket{\Phi_+}$.

Let us now focus on the study of the PPT region, resulting from the
intersection $\hat \Sigma^\Gamma\cap\Sigma$. As we mentioned, when
studying separability, one should characterize the expectation value
of the generators of the group with product vectors. For a vector
$\ket u\otimes \ket v$ one obtains that:
\begin{eqnarray}
&\langle\V\rangle&= \big|\langle u|v\rangle\big|^2 \nonumber\\
&=&\big|\bar u_0v_0+\bar u_1 v_1+\bar u_2 v_2+\bar u_3 v_3+
\dots+\bar u_{2n}v_{2n}\big|^2, \label{<F>} \nonumber \\
&\langle P_+^J\rangle&=\frac{1}{d}\big|u^TJv\big|^2\nonumber\\
&=&\frac{1}{d}\big|u_0v_1-u_1v_0+%u_2v_3-u_3v_2+
\dots+u_{2n-1}v_{2n}-u_{2n}v_{2n-1}\big|^2
. \nonumber\\
\label{<P+J>}
\end{eqnarray}
From these equations,
one easily sees that the first
extreme point from (\ref{points})
can be realized by e.g.
$u=(1/2)(-1,1,1,1,0,\dots)$ and $v=1/\sqrt 2(1,0,0,1,0,\dots)$, while
points $x_2,x_3$ can be obtained from
$u=1/\sqrt 2\big(|0\rangle\mp|1\rangle\big)$,
$v=1/\sqrt 2\big(|0\rangle+|1\rangle\big)$ respectively.
To show that only the set $\text{conv}\{x_0,x_1,x_2\}$
is separable we will employ the Breuer-Hall map (\ref{BH})
itself.
Note that the corresponding separable set
$\text{conv}\{x_0,x_1,x_2\}^\Gamma\subset\hat\Sigma\cap\Sigma^\Gamma$
is determined by the points with the same coordinates
as $x_0,x_1,x_2$ but in the $\langle P_+\rangle,
\langle \V^J\rangle$-plane (since e.g. $\tr(\varrho^\Gamma P_+^J)=1/d$
$\Leftrightarrow$$\tr(\varrho\V)=1$, etc).

For an arbitrary $SS$-invariant normalized state
$\varrho=\alpha\1+\beta\V+\gamma P_+^J$ it holds
$\tr_B\varrho=\big(d\alpha+\V+(1/d)\gamma\big)\1=\1/d$,
since $\tr\varrho=d^2\alpha+d\beta+\gamma=1$ and $\tr_BP_+^J=\tr_BP_+=\1/d$
as $J$ is unitary. Analogously, for an arbitrary $S\bar S$-invariant
state $\hat\varrho=\hat\alpha\1+\hat\beta\V^J+\hat\gamma P_+$,
$\tr_B\hat\varrho=\1/d$, since $\tr_B\V^J=\tr_B\V=\1$.
Hence, the no-detection
condition $\1\otimes\Lambda_{BH}(\varrho)\ge 0$ takes the same form
for both families:
\begin{equation}\label{BHviol}
\frac{1}{d}\1-\varrho-\big(\1\otimes J\big)
\varrho^\Gamma\big(\1\otimes J^\dagger\big)\ge 0.
\end{equation}
We multiply the above inequality by $P_+^J$ and $P_+$ respectively.
Noting that $P_+^J, P_+\ge 0$ and
$[\1\otimes\Lambda_{BH}(\varrho),P_+^J]=0=[\1\otimes\Lambda_{BH}(\hat\varrho),P_+]$,
we obtain that if a state is {\it not} detected by the Breuer-Hall map then:
\begin{eqnarray}
& &\tr(\varrho P_+^J)\le\frac{1-\tr(\varrho\V)}{d},\quad \text{or}\label{PPTBE1}\\
& &\tr(\hat\varrho P_+)\le\frac{1-\tr(\hat\varrho\V^J)}{d}\label{PPTBE2}
\end{eqnarray}
respectively. Equivalently, states breaking the above inequalities,
i.e. states lying above the line $\langle P_+^J\rangle
=\big(1-\langle\V\rangle\big)/d$, or above the line
$\langle P_+\rangle=\big(1-\langle\V^J\rangle\big)/d$ in the case
of $S\bar S$-invariant states, are detected by $\Lambda_{BH}$ and hence
entangled.

The set of PPT entangled $S\bar S$-invariant states is depicted in
Fig. \ref{figa1}. Note that when $d\to \infty$, $d$ even, the point
$x_3\to x_2$, cf. Eq. (\ref{points}), and the set of PPT bound
entangled states collapses. Since we expect that away from region
boundaries in Fig. \ref{figa1} the properties of  $S\bar
S$-invariant states are shared by the states in a small ball around
them, the collapse of the "volume" of the PPT states is to be
expected according to Ref. \cite{PPTBEinf}. From the previous
arguments (cf. remarks after Eq. (\ref{conv})) and Eq.
(\ref{PPTBE2}), the corresponding diagram for $SS$-invariant states
is identical, modulo the labels of the axes. This finishes our
analysis of unitary symplectic invariant states.


\begin{thebibliography}{99}
\bibitem{notepos} Throughout this paper, and for the sake of simplicity, we often name
as positive maps those maps that are positive but not completely positive.
\bibitem{note2} For a general unitary symplectic invariant operator
$A^\Gamma\ne A^J$, although it can happen (see Eq. (\ref{EG=EJ})).
\end{thebibliography}
\end{document}